\documentclass[a4paper,12pt]{article}
\usepackage{graphicx}

\begin{document}

%SHORTCUTS \input{../Macros/shortcuts}
%%%%%%%%%Fawzi's Short CUT: GENERAL%%%%%%%%%%%%%%%%%

\newcommand{\be}{\begin{equation}}
\newcommand{\beq}{\begin{equation}}
\newcommand{\eeq}{\end{equation}}
\newcommand{\ee}{\end{equation}}

\newcommand{\beqn}{\begin{eqnarray}}
\newcommand{\eeqn}{\end{eqnarray}}
\newcommand{\bea}{\begin{eqnarray}}
\newcommand{\ena}{\end{eqnarray}}
\newcommand{\ra}{\rightarrow}
\newcommand{\susy}{{{\cal SUSY}$\;$}}
\newcommand{\su}{$ SU(2) \times U(1)\,$}

\newcommand{\gag}{$\gamma \gamma$ }
\newcommand{\gagt}{\gamma \gamma }
\newcommand{\gam}{\gamma \gamma }
\def\W{{\mbox{\boldmath $W$}}}
\def\B{{\mbox{\boldmath $B$}}}
\def\V{{\mbox{\boldmath $V$}}}
\newcommand{\np}{Nucl.\,Phys.\,}
\newcommand{\pl}{Phys.\,Lett.\,}
\newcommand{\pr}{Phys.\,Rev.\,}
\newcommand{\prl}{Phys.\,Rev.\,Lett.\,}
\newcommand{\prep}{Phys.\,Rep.\,}
\newcommand{\zp}{Z.\,Phys.\,}
\newcommand{\sovjnp}{{\em Sov.\ J.\ Nucl.\ Phys.\ }}
\newcommand{\nuclinst}{{\em Nucl.\ Instrum.\ Meth.\ }}
\newcommand{\annp}{{\em Ann.\ Phys.\ }}
\newcommand{\intjmp}{{\em Int.\ J.\ of Mod.\  Phys.\ }}

%GENERAL
\newcommand{\eps}{\epsilon}
\newcommand{\mw}{M_{W}}
\newcommand{\mww}{M_{W}^{2}}
\newcommand{\mwmw}{M_{W}^{2}}
\newcommand{\mhmh}{M_{H}^2}
\newcommand{\mz}{M_{Z}}
\newcommand{\mzz}{M_{Z}^{2}}

\newcommand{\cw}{\cos\theta_W}
\newcommand{\sw}{\sin\theta_W}
\newcommand{\tw}{\tan\theta_W}
\def\cww{\cos^2\theta_W}
\def\sww{\sin^2\theta_W}
\def\tww{\tan^2\theta_W}
\def\stw{s_{2W}}

\newcommand{\epm}{$e^{+} e^{-}\;$}
\newcommand{\epemt}{$e^{+} e^{-}\;$}
\newcommand{\epem}{e^{+} e^{-}\;}
\newcommand{\ememt}{$e^{-} e^{-}\;$}
\newcommand{\emem}{e^{-} e^{-}\;}

\newcommand{\lra}{\leftrightarrow}
\newcommand{\tr}{{\rm Tr}}
\def\ls1{{\not l}_1}
\newcommand{\cms}{centre-of-mass\hspace*{.1cm}}

%W Physics and anomalous

\newcommand{\dkg}{\Delta \kappa_{\gamma}}
\newcommand{\dkz}{\Delta \kappa_{Z}}
\newcommand{\dz}{\delta_{Z}}
\newcommand{\dgz}{\Delta g^{1}_{Z}}
\newcommand{\dgzt}{$\Delta g^{1}_{Z}\;$}
\newcommand{\la}{\lambda}
\newcommand{\lag}{\lambda_{\gamma}}
\newcommand{\lambdae}{\lambda_{e}}
\newcommand{\laz}{\lambda_{Z}}
\newcommand{\lnl}{L_{9L}}
\newcommand{\lnr}{L_{9R}}
\newcommand{\lt}{L_{10}}
\newcommand{\lu}{L_{1}}
\newcommand{\ld}{L_{2}}
\newcommand{\eeww}{e^{+} e^{-} \ra W^+ W^- \;}
\newcommand{\eewwt}{$e^{+} e^{-} \ra W^+ W^- \;$}
\newcommand{\epemww}{e^{+} e^{-} \ra W^+ W^- }
\newcommand{\epemwwt}{$e^{+} e^{-} \ra W^+ W^- \;$}
\newcommand{\eennhht}{$e^{+} e^{-} \ra \nu_e \bar \nu_e HH\;$}
\newcommand{\eennhh}{e^{+} e^{-} \ra \nu_e \bar \nu_e HH\;}
\newcommand{\ppwg}{p p \ra W \gamma}
\newcommand{\wwhh}{W^+ W^- \ra HH\;}
\newcommand{\wwhht}{$W^+ W^- \ra HH\;$}
\newcommand{\ppwz}{pp \ra W Z}
\newcommand{\ppwgt}{$p p \ra W \gamma \;$}
\newcommand{\ppwzt}{$pp \ra W Z \;$}
\newcommand{\gamgamt}{$\gamma \gamma \;$}
\newcommand{\gamgam}{\gamma \gamma \;}
\newcommand{\egamt}{$e \gamma \;$}
\newcommand{\egam}{e \gamma \;}
\newcommand{\gamgamwwt}{$\gamma \gamma \ra W^+ W^- \;$}
\newcommand{\gamgamwwht}{$\gamma \gamma \ra W^+ W^- H \;$}
\newcommand{\gamgamwwh}{\gamma \gamma \ra W^+ W^- H \;}
\newcommand{\gamgamwwhht}{$\gamma \gamma \ra W^+ W^- H H\;$}
\newcommand{\gamgamwwhh}{\gamma \gamma \ra W^+ W^- H H\;}
\newcommand{\ggww}{\gamma \gamma \ra W^+ W^-}
\newcommand{\ggwwt}{$\gamma \gamma \ra W^+ W^- \;$}
\newcommand{\ggwwht}{$\gamma \gamma \ra W^+ W^- H \;$}
\newcommand{\ggwwh}{\gamma \gamma \ra W^+ W^- H \;}
\newcommand{\ggwwhht}{$\gamma \gamma \ra W^+ W^- H H\;$}
\newcommand{\ggwwhh}{\gamma \gamma \ra W^+ W^- H H\;}
\newcommand{\ggwwz}{\gamma \gamma \ra W^+ W^- Z\;}
\newcommand{\ggwwzt}{$\gamma \gamma \ra W^+ W^- Z\;$}

\newcommand{\ptu}{p_{1\bot}}
\newcommand{\vecptu}{\vec{p}_{1\bot}}
\newcommand{\ptd}{p_{2\bot}}
\newcommand{\vecptd}{\vec{p}_{2\bot}}
\newcommand{\ie}{{\em i.e.}}
\newcommand{\cm}{{{\cal M}}}
\newcommand{\cl}{{{\cal L}}}
\newcommand{\cd}{{{\cal D}}}
\newcommand{\cv}{{{\cal V}}}
\def\slashc{c\kern -.400em {/}}
\def\slashp{p\kern -.400em {/}}
\def\slashL{L\kern -.450em {/}}
\def\slashcl{\cl\kern -.600em {/}}
\def\Ww{{\mbox{\boldmath $W$}}}
\def\B{{\mbox{\boldmath $B$}}}
\def\noi{\noindent}
\def\nn{\noindent}
\def\sm{${\cal{S}} {\cal{M}}\;$}
\def\smn{${\cal{S}} {\cal{M}}$}
\def\nph{${\cal{N}} {\cal{P}}\;$}
\def\sb{$ {\cal{S}}  {\cal{B}}\;$}
\def\ssb{${\cal{S}} {\cal{S}}  {\cal{B}}\;$}
\def\ssbe{{\cal{S}} {\cal{S}}  {\cal{B}}}
\def\cviol{${\cal{C}}\;$}
\def\pviol{${\cal{P}}\;$}
\def\cpviol{${\cal{C}} {\cal{P}}\;$}

\newcommand{\lgg}{\lambda_1\lambda_2}
\newcommand{\lww}{\lambda_3\lambda_4}
\newcommand{\ppin}{ P^+_{12}}
\newcommand{\pmin}{ P^-_{12}}
\newcommand{\ppout}{ P^+_{34}}
\newcommand{\pmout}{ P^-_{34}}
\newcommand{\sinsq}{\sin^2\theta}
\newcommand{\cossq}{\cos^2\theta}
\newcommand{\yt}{y_\theta}
\newcommand{\hppll}{++;00}
\newcommand{\hpmll}{+-;00}
\newcommand{\hpplt}{++;\lambda_30}
\newcommand{\hpmlt}{+-;\lambda_30}
\newcommand{\hpptt}{++;\lambda_3\lambda_4}
\newcommand{\hpmtt}{+-;\lambda_3\lambda_4}
\newcommand{\dk}{\Delta\kappa}
\newcommand{\klam}{\Delta\kappa \lambda_\gamma }
\newcommand{\kac}{\Delta\kappa^2 }
\newcommand{\lac}{\lambda_\gamma^2 }
\def\gamgamtzz{$\gamma \gamma \ra ZZ \;$}
\def\gamgamtww{$\gamma \gamma \ra W^+ W^-\;$}
\def\gamgamtwwe{\gamma \gamma \ra W^+ W^-}

%SUSY DEFINITIONS \input{../Macros/susydef}
\def\sinb{\sin\beta}
\def\cosb{\cos\beta}
\def\sinbb{s_ {2\beta}}
\def\cosbb{c_{2 \beta}}
\def\tgb{\tan \beta}
\def\tgbt{$\tan \beta\;\;$}
\def\tgbsq{\tan^2 \beta}
\def\tgbsqt{$\tan^2 \beta\;$}
\def\sinal{\sin\alpha}
\def\cosal{\cos\alpha}
%%%short notation of angles%%%%%
\def\sb{s_\beta}
\def\cb{c_\beta}
\def\tb{t_\beta}
\def\ttb{t_{2 \beta}}
\def\sa{s_\alpha}
\def\ca{c_\alpha}
\def\ta{t_\alpha}
\def\stb{s_{2\beta}}
\def\ctb{c_{2\beta}}
\def\sbb{s_ {2\beta}}
\def\cbb{c_{2 \beta}}
\def\sta{s_{2\alpha}}
\def\cta{c_{2\alpha}}
\def\sbma{s_{\beta-\alpha}}
\def\cbma{c_{\beta-\alpha}}
\def\sbpa{s_{\beta+\alpha}}
\def\cbpa{c_{\beta+\alpha}}
\def\lone{\lambda_1}
\def\ltwo{\lambda_2}
\def\lthree{\lambda_3}
\def\lfour{\lambda_4}
\def\lfive{\lambda_5}
\def\lsix{\lambda_6}
\def\lseven{\lambda_7}
%%%%%%%%%%%%%%%%%%%%%%%%5
\def\stop{\tilde{t}}
\def\sto{\tilde{t}_1}
\def\stt{\tilde{t}_2}
\def\stl{\tilde{t}_L}
\def\str{\tilde{t}_R}
\def\msto{m_{\sto}}
\def\mstosq{m_{\sto}^2}
\def\mstt{m_{\stt}}
\def\msttsq{m_{\stt}^2}
\def\mt{m_t}
\def\mtsq{m_t^2}
\def\sint{\sin\theta_{\stop}}
\def\sintt{\sin 2\theta_{\stop}}
\def\cost{\cos\theta_{\stop}}
\def\sintsq{\sin^2\theta_{\stop}}
\def\costsq{\cos^2\theta_{\stop}}
\def\mqtt{\M_{\tilde{Q}_3}^2}
\def\mutt{\M_{\tilde{U}_{3R}}^2}
%%%%%%%%%%%%%%%%%%%%%
\def\sbottom{\tilde{b}}
\def\sbo{\tilde{b}_1}
\def\sbt{\tilde{b}_2}
\def\sbl{\tilde{b}_L}
\def\sbr{\tilde{b}_R}
\def\msbo{m_{\sbo}}
\def\msbosq{m_{\sbo}^2}
\def\msbt{m_{\sbt}}
\def\msbtsq{m_{\sbt}^2}
\def\mt{m_t}
\def\mtsq{m_t^2}
%%%%%%%%%%%%%%%%%%%%%
\def\selectron{\tilde{e}}
\def\seo{\tilde{e}_1}
\def\set{\tilde{e}_2}
\def\sel{\tilde{e}_L}
\def\ser{\tilde{e}_R}
\def\mseo{m_{\seo}}
\def\mseosq{m_{\seo}^2}
\def\mset{m_{\set}}
\def\msetsq{m_{\set}^2}
\def\msel{m_{\sel}}
\def\mser{m_{\ser}}
\def\me{m_e}
\def\mesq{m_e^2}
%%%%%%%%%%%%%%%%%%%%%
\def\snu{\tilde{\nu}}
\def\snue{\tilde{\nu_e}}
\def\set{\tilde{e}_2}
\def\snul{\tilde{\nu}_L}
\def\msnue{m_{\snue}}
\def\msnuesq{m_{\snue}^2}
%%%%%%%%%%%%%%%%%%%%%
\def\smuon{\tilde{\mu}}
\def\smul{\tilde{\mu}_L}
\def\smur{\tilde{\mu}_R}
\def\msmul{m_{\smul}}
\def\msmulsq{m_{\smul}^2}
\def\msmur{m_{\smur}}
\def\msmursq{m_{\smur}^2}
%%%%%%%%%%%%%%%%%%%%%%%%%%
\def\stau{\tilde{\tau}}
\def\stauo{\tilde{\tau}_1}
\def\staut{\tilde{\tau}_2}
\def\staul{\tilde{\tau}_L}
\def\staur{\tilde{\tau}_R}
\def\mstauo{m_{\stauo}}
\def\mstauosq{m_{\stauo}^2}
\def\mstaut{m_{\staut}}
\def\mstautsq{m_{\staut}^2}
\def\mtau{m_\tau}
\def\mtausq{m_\tau^2}
%%%%%%%%%%%%%%%%%%%%%%%%%%%%%%
\def\gluino{\tilde{g}}
\def\mgluino{m_{\tilde{g}}}
\def\mchi{m_\chi^+}
\def\neuto{\tilde{\chi}_1^0}
\def\mneuto{m_{\tilde{\chi}_1^0}}
\def\neutt{\tilde{\chi}_2^0}
\def\mneutt{m_{\tilde{\chi}_2^0}}
\def\neutth{\tilde{\chi}_3^0}
\def\mneutth{m_{\tilde{\chi}_3^0}}
\def\neutf{\tilde{\chi}_4^0}
\def\mneutf{m_{\tilde{\chi}_4^0}}
\def\chargop{\tilde{\chi}_1^+}
\def\mchargo{m_{\tilde{\chi}_1^+}}
\def\chargtp{\tilde{\chi}_2^+}
\def\mchargt{m_{\tilde{\chi}_2^+}}
\def\chargom{\tilde{\chi}_1^-}
\def\chargtm{\tilde{\chi}_2^-}
\def\bino{\tilde{b}}
\def\wino{\tilde{w}}
\def\photino{\tilde{\gamma}}
\def\zino{tilde{z}}
%%%%%%%%%%%%%%%%%%%%%%%%%%%%%%%%%
\def\sdowno{\tilde{d}_1}
\def\sdownt{\tilde{d}_2}
\def\sdownl{\tilde{d}_L}
\def\sdownr{\tilde{d}_R}
\def\supo{\tilde{u}_1}
\def\supt{\tilde{u}_2}
\def\supl{\tilde{u}_L}
\def\supr{\tilde{u}_R}
%%%%%%%%%%Higgses masses%%%%%%%%%%%%
\def\mh{m_h}
\def\mht{m_h^2}
\def\MH{M_H}
\def\MHt{M_H^2}
\def\MA{M_A}
\def\MAt{M_A^2}
\def\MHp{M_H^+}
\def\MHm{M_H^-}
%%%%%%%%%%NEEDED FOR rge%%%%%%%%%%%%
\def\mqt{\M_{\tilde{Q}_3}}
\def\mut{\M_{\tilde{U}_{3R}}}
\def\mqtz{\M_{\tilde{Q}_3(0)}}
\def\mutz{\M_{\tilde{U}_{3R}(0)}}
\def\mqtzt{\M_{\tilde{Q}_3^2(0)}}
\def\mutzt{\M_{\tilde{U}_{3R}^2(0)}}

\begin{titlepage}
%***************************************************
%********************LAPTH HEADER****************
%\input{entete}
%****************************************************
\def\baselinestretch{1.2}
\topmargin     -0.25in
\vspace*{\fill}
\begin{center}
{\large {\bf  Measurements of the SUSY Higgs self-couplings and
the reconstruction of the Higgs potential}}

\vspace*{0.5cm}
%\fill}

\begin{tabular}[t]{c}

{\bf  F.~Boudjema and A.~Semenov }
 \\
\\
\\
{\it  Laboratoire de Physique Th\'eorique} {\large
LAPTH}$^\dagger$
\\
 {\it Chemin de Bellevue, B.P. 110, F-74941 Annecy-le-Vieux,
Cedex, France.}\\
\end{tabular}
\end{center}

\centerline{ {\bf Abstract} } \baselineskip=14pt \noindent
%%%%%%%%%%%%%%%%%%%%%%%%%%%%%%%%%%%%%%%%%%%%%%%%%%%%%%%%%%%%%%%%%%%%%
 {\small We address the issue of the reconstruction of the scalar
 potential of a two-Higgs doublet model having in mind that of the MSSM.
 We first consider the general CP conserving dim-4 effective
 potential. To fully reconstruct this potential, we show that even if all
 the Higgs masses and their couplings to the standard model particles are
 measured one needs not only to measure certain trilinear Higgs self-couplings but
 some of the quartic couplings as well. We  also advocate  expressing the Higgs self
 couplings in the mass basis. We show explicitly, that in
 the so-called decoupling limit, the most easily accessible Higgs
 self-couplings are given in terms of the Higgs mass while all
 other dependencies on the parameters of the general effective
 potential are screened. This helps also easily explain how, in the
 MSSM, the largest radiative corrections which affect these self couplings
 are reabsorbed by using the corrected Higgs mass. We
 also extend our analysis to higher order operators in the
 effective Higgs potential. While the above screening properties
 do not hold, we argue that these effects must be small and may
 not be measured considering the foreseen poor experimental precision in the
 extraction of the SUSY Higgs self-couplings.
}
%%%%%%%%%%%%%%%%%%%%%%%%%%%%%%%%%%%%%%%%%%%%%%%%%%%%%%%%%%%%%%%%%%%%%
\vspace*{\fill}

%\hrulefill\ $\; \; \; \; \; \; \; \; \; \; \; \; \; \; \; \; \; \; \; \;\;$
%\hspace*{3.5cm}\\

\vspace*{0.1cm} \rightline{LAPTH-880/01}
%\rightline{xxxx/x-98/xxx} \rightline{hep-ph/yymmddd}
%\today \\
\rightline{{\large  December 2001}}

$^\dagger${\small URA 14-36 du CNRS, associ\'ee  \`a
l'Universit\'e de Savoie.} \normalsize
\end{titlepage}
%\baselineskip=18pt

%*********************** counters ********************
\setcounter{section}{0} \setcounter{subsection}{0}
\setcounter{equation}{0}
\def\thesubsection {\thesection.\arabic{subsection}}
\def\theequation{\thesection.\arabic{equation}}
\section{Introduction}
The most important issue at the upcoming colliders is the
elucidation of the mechanism of symmetry breaking and the hunt for
the Higgs. Within the Standard Model, \sm, there is a strong
indirect evidence that the latter might be light. But at the same
time within the \sm such a light Higgs poses the problem of
naturalness. Supersymmetry, SUSY, solves this problem and for a
large array of models predicts a light Higgs in accordance with
the present precision data. The task of the next colliders will
therefore be not so much the discovery of the (lightest) Higgs but
a careful study of the properties of the Higgs system since this
will be an ideal window on the mechanism of (super)symmetry
breaking. Many state-of-the-art  studies have analysed the
couplings of the lightest Higgs to fermions and to gauge bosons.
Most useful conclusions are in the context of a next linear
collider (LC) \footnote{A Higgs factory at a muon facility gives
astounding results\cite{Higgsatmucoll}, however many technical
problems need to be solved before the design of such a facility.},
for a summary see \cite{BattagliaDesch,Teslaphysics}. One can for
example discriminate between a light Higgs within the \sm and one
within a supersymmetric model. Precision studies on other
supersymmetric particles that may be produced at these colliders
can nicely complement these studies. However one key ingredient
that still requires further studies and simulations concerns the
important aspect of the Higgs potential. Not only because this
triggers electroweak symmetry breaking but also because
supersymmetry breaking is also encoded in this potential. Some
studies\cite{trilinearHiggsMSSMee,Panditatrilinear,Muhlthesis,trilinearHiggsMSSMlhc,Lafayehhh}
have addressed the issue of the measurements of some of the Higgs
self-couplings within minimal SUSY. These studies have been rather
purely phenomenological studies in the sense that one has, within
the minimal supersymmetric model (MSSM), quantified various cross
sections for double Higgs (and for some triple) Higgs production
at a high energy collider. From these one has derived a
sensitivity on some individual Higgs self-couplings by varying the
strength of these couplings independently, while fixing the Higgs
mass spectrum. We would however expect that a deviation in one of
the Higgs self-couplings should  not only affect other Higgs
self-couplings but also affect the Higgs mass spectrum. Moreover,
one would like to see how a particular deviation in the Higgs
self-couplings relates to the fundamental parameters of the Higgs
potential and also what order of magnitude should one expect from
these deviations. This can most efficiently be addressed through
an effective potential approach and would be similar to what has
been applied in the measurements of the trilinear\cite{HPZH} or
even quadrilinear\cite{quadrilinearw} self-couplings of the weak
vector bosons. Within the one Higgs doublet of the \sm a general
parameterisation of the Higgs self-couplings has been
given\cite{Jikia3H,ChopinHiggs} and its effects at the colliders
studied\cite{Jikia3H,ChopinHiggs,selfHiggsSM}. As for the case of
SUSY where one needs two Higgs doublets such a study is missing
although a leading order parameterisation has been
known\cite{HaberHempfling}. The aim of this paper is to fill this
gap. As we will see this parameterisation of the scalar potential
is important, not only it can embody through an effective
potential many of the well known radiative
corrections\cite{brignole1,brignole2,brignole3,RCmhtrilinear,mhSUSYcalc}
but it will also make clear the link between what can be learned
about the Higgs potential by measuring the Higgs masses and what
additional, if any, information can be gained if one can study the
self-couplings. This will also show that the couplings of the
charged Higgs to the neutral Higgses can embody the same
information as some of the trilinear neutral Higgs couplings. The
measurement of the Higgs self-couplings involving the charged
Higgses have, as far as we know, never been addressed although a
calculation of charged Higgs pair production at the LHC has been
made\cite{Muhlthesis,hpluslhc}. Our findings also help understand
why the radiative corrections to the Higgs self-couplings of the
lightest Higgs in the MSSM though substantial ( as are those to
the Higgs mass) become tiny when expressed in terms of the
lightest Higgs mass\cite{Holliksusyselfcouplings}. With the five
$(h,H,A,H^\pm)$ Higgses of the MSSM one has 8 possible Higgs
trilinear self-couplings. One would then think that,  together
with the measurements of the Higgs masses and their couplings to
ordinary matter the measurement of these trilinear self-couplings
would allow a full reconstruction of the leading order dim-4
effective Lagrangian describing the Higgs potential. Even in the
most optimistic scenarios where all the Higgses are light, we find
that a full reconstruction requires the measurements of some
quartic couplings which are extremely difficult, if not
impossible\cite{hhhbattagliaboos}, to measure even at the linear
collider. In the decoupling regime\cite{HaberHempfling}where only
one of the Higgses is light and with properties very much
resembling those of the \sm, the effective  parameters of the
Higgs potential will be screened and thus extremely difficult to
measure.  One can, of course,
 entertain that new physics affecting the Higgs potential appears
 as higher dimensional operators, dim-6, in which case the above
 ``screening" effects are not operative. However one expects these
 effects to be too small to be measured considering the expected
 accuracy, no better than $\sim 10\%$, at which the Higgs self-couplings are to be probed
 (for a nice review see\cite{BattagliaDesch,Teslaphysics,hhhbattagliaboos}).

\setcounter{equation}{0}
\section{Leading order parameterisation of the Higgs potential and
the Higgs self-couplings}
 When trying to parameterise the effects
of some new physics on the properties of {\em familiar} particles,
the effective Lagrangian is most useful. One uses all the known
symmetries of the model and then writes the tower of operators
according to their dimensions. One expects thus that the allowed
lowest
 dimension operators have most impact on the low energy
observables, higher order operators being suppressed due to the
large mass scale needed for their parameterisation. Therefore, for
the minimal SUSY the lowest dimension operator is of dim-4. For
the MSSM one needs two Higgs doublets, $H_1$ and $H_2$ with
opposite hypercharge, $Y=\mp 1$ respectively. They may be written
as

\beqn
\label{doublets}
 H_1=\left(\begin{array}{c}v_1
+\frac{1}{\sqrt{2}}(h_1+i\varphi_1^0) \\
\varphi_1^- \end{array} \right) \;\;\;\;
H_2=\left(\begin{array}{c} \varphi_2^+ \\
v_2 +\frac{1}{\sqrt{2}}(h_2+i\varphi_2^0)
\end{array}\right)
\eeqn

Requiring \cpviol conservation one can write
following\cite{HaberHempfling},

\begin{eqnarray}
\label{potential4}
V_{eff} & = & (m_{1}^2+\mu^2) |H_1|^2 +
(m_{2}^2+\mu^2) |H_2|^2
              - [m_{12}^2 (\epsilon H_1 H_2) + h.c. ] \nonumber \\
        &   & + \frac{1}{2}\big[\frac{1}{4}(g^2+g'^2) + \lambda_1\big]
                            (|H_1|^2)^2
              +\frac{1}{2}\big[\frac{1}{4}(g^2+g'^2) + \lambda_2\big]
                            (|H_2|^2)^2 \nonumber \\
        &  &  +\big[\frac{1}{4}(g^2-g'^2) + \lambda_3\big]
                            |H_1|^2 |H_2|^2
              +\big[-\frac{1}{2}g^2 + \lambda_4\big]
(\epsilon H_1 H_2)(\epsilon H_1^* H_2^*)\nonumber\\
        & &   +\big(\frac{\lambda_5}{2} (\epsilon H_1 H_2)^2
                   +\big[ \lambda_6 |H_1|^2 + \lambda_7 |H_2|^2\big]
                         (\epsilon H_1 H_2) + h.c. \big)
\end{eqnarray}

$\epsilon$ is the antisymmetric matrix with $\epsilon_{12}=-1$.
$g$ and $g'$ are the SU(2) and U(1) gauge couplings. $\mu$ is a
supersymmetry preserving mass term. The case with all $\lambda_i$
being zero corresponds to the original MSSM potential at
tree-level. Moreover, exact supersymmetry  imposes that
$m_1^2=m_2^2=m_{12}^2=0$ so that no electroweak symmetry breaking
ensues. $m_1,m_2$ and $m_{12}$ are thus essential for electroweak
symmetry breaking and  encode also supersymmetry breaking. These
dimensionful quantities are soft SUSY breaking parameters. As for
the $\lambda$'s, practically all analyses of the Higgs
phenomenology have only viewed them as ``soft" terms originating
from higher orders loop effects. As known these loop effects can
be substantial as they are enhanced by large Yukawa couplings
(corrections are quartic in the top mass) and have kept the MSSM
alive, see \cite{CarenaWagner_Higgs_Approx1} for example. However
it may well be that models of supersymmetry breaking can provide a
{\em direct} contribution to these parameters, for instance
through non-renormalisable operators in the K\"ahler potential.
Technically these contributions would be deemed {\em hard}. It has
however been stressed
recently\cite{Polonskihardterms,SPMartinhardterms} that if these
parameters are related to the source of soft SUSY breaking then
they would not destabilise the scalar potential and would evade
the ``un-natural" quadratic divergence problem, thus leading to a
viable model. In such circumstances such terms may lead to the
lightest SUSY Higgs with a mass much in excess of $150$GeV and
with ``no un-naturalessness" dilemma. Therefore, the
reconstruction of the Higgs potential is crucial. In a different
context, supersymmetric models with a warped fifth
dimension\cite{Casashiggswarped}, it was shown that some of the
quartic couplings (apart from the usual ones of gauge origin) can
even be supersymmetric and originate from a non-minimal K\"ahler
potential. Supersymmetry breaking terms also contribute to the
latter as well as to the quadratic terms. However to obtain a
satisfactory electroweak breaking in this scheme constrains the
parameters such that $\tb=1$ is picked up (see below the
definition of $\tb$).\\
\noi Of course, the approach we follow and the results we obtain
can be made to apply directly to a general 2-Higgs doublet
model(2HDM) , one only needs to switch the gauge couplings
contributions off in the potential Eq.~\ref{potential4}. This said
when we investigate precision measurements extracted from the
Higgs bosons to fermions one should stick to a model whose
characteristics are close to the MSSM. The 2HDM we have in mind is
the so-called type-II, in the terminology of \cite{HHG}, where the
down-type quarks and leptons couple to $H_1$ and the up-type
quarks to $H_2$ as in the MSSM. However in most studies of the
2HDM $\lambda_{6,7}$ are not considered on the basis that they may
induce too large FCNC.\\
\noi This brings us to the issue concerning the order of magnitude
for the various $\lambda_i$ in a general suspersymmetric context.
For the conventional MSSM, one effectively gets a Yukawa enhanced
contribution, starting at one loop, which affects all  seven
paramaters\cite{CarenaWagner_Higgs_Approx1}. The largest
contribution for moderate $\tgb$ stems from $\lambda_2$ and is of
order $.1$. In \cite{Polonskihardterms} where ``natural hard"
terms are discussed, some orders of magnitude for the $\lambda_i$
are given based on how and at what scale SUSY breaking is
transmitted. Values as high as $1.$ could be entertained. Such
values ($< 4\pi$) are still perturbative. In the warped fifth
dimension model\cite{Casashiggswarped} it is interesting to note
that all $\lambda_i$ but $\lambda_5$ get a SUSY conserving
contribution in addition to some SUSY breaking contribution, while
$\lambda_{5}$ is of purely SUSY breaking origin. All these
contribution disappear in the limit of an infinitely large
effective fundamental scale (scale {\it in lieu} of $M_{\rm
Planck}$), which is directly related to the warp factor. The
effective couplings can be large enough, in fact so large that the
authors estimate a lightest Higgs mass of order $700$GeV as a
possibility while the couplings are still perturbative up to the
cut-off scale. Note however that the cut-off scale is identified
with the ``low" fundamental scale as compared to the usual Planck
scale that is usually used to set the upper bound on the quartic
coupling and hence the mass of the Higgs. Moreover  it is argued
that the model is safe as regards FCNC \cite{Casashiggswarped} .
\\ \noi Some bounds, though not so strong, exist.
In all its generality, if we allow some combinations of couplings
within the range $-1< \lambda_i < 1$, independently of $\tgb$, can
lead to too low values of the Higgs masses (even negative squared
masses and problems with vacuum stability can occur). Furthermore
to constrain the parameter space some authors have imposed vacuum
stability of the potential and perturbativity of the couplings up
to high scales\cite{Kanemura2hdm} as well as   tree-level
unitarity constraints of the elastic scattering of the
Higgses\cite{unitarity2hdm} in the 2HDM. Limits from $\Delta
\rho$\cite{deltarho2hdm} can also be quite useful, although they
are model dependent if for instance the stop contribution to
$\Delta \rho$ plays a role. Nonetheless all of these requirements
still leave a large parameter space that can be quite drastically
reduced once the Higgs masses and their couplings are directly
measured.

We will therefore follow a general approach based on the potential
\ref{potential4}, assuming the quadratic terms ($m_{1,2,12}$) to
fulfill the usual conditions for a stable minimum with
non-vanishing vacuum expectation values. The minimisation of the
potential, and the absence of tadpoles, imposes the following
constraint on the ``soft" SUSY parameters $m_1, m_2$:

\begin{eqnarray}
m_{1}^2 & = & -m_{12}^2 t_\beta -\mu^2-M_Z^2c_{2\beta}/2 + v^2
     (-\lambda_1c_\beta^2 - \lambda_3 s_\beta^2-\lambda_4s_\beta^2
     -\lambda_5s_\beta^2  \nonumber \\
     &+&  3\lambda_6c_\beta s_\beta+\lambda_7s_\beta^3/c_\beta)\\
m_{2}^2 & = & -m_{12}^2 / t_\beta -\mu^2+M_Z^2c_{2\beta}/2 + v^2
     (-\lambda_2s_\beta^2 - \lambda_3 c_\beta^2-\lambda_4c_\beta^2
     -\lambda_5c_\beta^2  \nonumber \\
     &+&  \lambda_6c_\beta^3/s_\beta+3\lambda_7c_\beta s_\beta)
\end{eqnarray}
where $v^2=v_1^2+v_2^2=2M_W^2/g^2$, $t_\beta=tan\,\beta=v_2/v_1$,
$s_\beta=\sin\beta$ and so on.

Then, the parameter $m_{12}^2$ can be fixed if we choose $M_A$,
the pseudo-scalar Higgs mass, as an  independent variable:

\beqn
m_{12}^2 = - c_\beta
s_\beta(M_A^2+v^2(2\lambda_5-\lambda_6/t_\beta
-\lambda_7t_\beta)).
\eeqn

\subsection{The Higgs masses}
At this stage, beside $M_A$ and $t_\beta$, there are  seven
independent parameters. Luckily some of these enter the
expressions of the Higgs masses and the couplings to fermions and
vector bosons.

The mass of the charged Higgs boson reads as
\beqn
M_{H^\pm}^2 = M_A^2+M_W^2-v^2(\lambda_4-\lambda_5).
\eeqn
This already shows that a  measurement of $M_{H^\pm}$ and $M_A$
can put a limit on the combination $(\lambda_4-\lambda_5)$. A
dedicated study addressing this particular issue at the LHC to
differentiate between the MSSM and a general 2HDM has very
recently appeared\cite{hplusyuanlhc}.

 The two CP-even Higgs states are determined by
the mixing angle $\alpha$. $h$ and H will denote the  \cpviol even
Higgs scalars defined through

\begin{eqnarray}
\label{tanadef}
 tan\,2\alpha & = &
\frac{M_A^2s_{2\beta}+M_Z^2s_{2\beta}-
                2v^2(s_{2\beta}(\lambda_3+\lambda_4)
                -2c_\beta^2\lambda_6-2s_\beta^2\lambda_7)   }{
                     M_A^2c_{2\beta}-M_Z^2c_{2\beta}-
                2v^2(\lambda_1c_\beta^2-\lambda_2s_\beta^2
                -\lambda_5c_{2\beta}-(\lambda_6-\lambda_7)s_{2\beta})  }\\[3mm]
                \label{mHdef}
M_H^2 & = & M_Z^2c_{\alpha+\beta}^2+M_A^2s_{\alpha-\beta}^2
\nonumber\\
      &   &  +2v^2(
             \lambda_1c_\alpha^2c_\beta^2+\lambda_2s_\alpha^2s_\beta^2
             +2(\lambda_3+\lambda_4)c_\alpha c_\beta s_\alpha s_\beta
             +\lambda_5(c_\alpha^2 s_\beta^2+s_\alpha^2 c_\beta^2) \nonumber \\
     &-&  2s_{\alpha+\beta}(\lambda_6c_\alpha c_\beta+\lambda_7s_\alpha
          s_\beta))\\
          \label{mhdef}
M_h^2 & = & M_Z^2s_{\alpha+\beta}^2+M_A^2c_{\alpha-\beta}^2 \nonumber\\
&   &  +2v^2(
             \lambda_1s_\alpha^2c_\beta^2+\lambda_2c_\alpha^2s_\beta^2
             -2(\lambda_3+\lambda_4)c_\alpha c_\beta s_\alpha s_\beta
             +\lambda_5(c_\alpha^2 c_\beta^2+s_\alpha^2 s_\beta^2) \nonumber \\
     &+& 2c_{\alpha+\beta}(\lambda_6s_\alpha c_\beta-\lambda_7c_\alpha
          s_\beta))
\end{eqnarray}

With $-\pi/2\leq \alpha \leq \pi/2$\footnote{$tan\,2\alpha$, does
not fully define $\alpha$, we should have given both $s_{2\alpha}$
and $c_{2\alpha}$.}, $h$ ($H$) defines the lightest (heaviest)
\cpviol even Higgs mass. The decoupling
limit\cite{Haber_decoupling} is usually defined as $M_A \gg M_Z$,
we will extend this to mean $M_A \gg v$ (with the $\lambda_{1-7}$
never exceeding ${\cal{O}}(1)$). Having the decoupling limit in
mind it is very instructive and useful to express the dependence
in the mixing angle $\alpha$ through $\cbma$ and $\sbma$ since
these two quantities are a direct measure of  the couplings of $h$
(and $H$) to vector bosons and to fermions. In units of the \sm
Higgs couplings, the couplings to vector bosons are
\beqn
g_{hVV,HVV}=\sbma,\cbma \nonumber
\eeqn
while for instance

\beqn
g_{hb\bar b}=-\sa/\cb=\sbma -\tb \cbma
\eeqn
\noi Therefore, especially if $\tb$ has been identified from other
SUSY processes, these combinations could be easily extracted once
the light Higgs has been produced at the linear collider, and
after allowing for some (important) QCD and in some cases model
dependent vertex corrections. The couplings of $H$ can be easily
translated from those of $h$ by the substitution $h \ra H, M_h \ra
M_H, \sbma \ra \cbma, \cbma \ra -\sbma$.  Moreover $\cbma$ is a
very good measure of decoupling since in this limit $\cbma \sim
1/M_A^2$. To wit,

\beqn
& &M_A^2 \cbma \sbma \sim M_A^2 \cbma \ra \\
 && \stb \ctb \left\{ M_Z^2 -
v^2 \left(\lthree+\lfour+\lfive +(\lsix-\lseven)\ttb-\lone
\frac{\cb^2}{\ctb}+\ltwo\frac{\sb^2}{\ctb}-\frac{\lsix}{\tb}-\lseven
\tb\right) \right\} \nonumber
\eeqn

It is important to keep in mind for later reference that in the
decoupling limit we have:
\beqn
\cbma\; , \; (M_A^2-M_H^2) \ra {{\cal O}}(1/M_A^2)
\eeqn

In the decoupling limit, the lightest Higgs mass writes as

\beqn
\label{mhdecoupling}
 M_h^2 \ra M_Z^2 \cbb^2 +2v^2 \left\{
\lambda_1 \cb^4+\lambda_2 \sb^4 +\sbb \big(
(\lambda_3+\lambda_4+\lambda_5)\sb \cb
-(\lambda_6+\lambda_7) -(\lambda_6-\lambda_7)\cbb \big) \right\}\nonumber \\
\eeqn

These simple considerations already show that if some of the
$\lambda_{1-7}$ are not too tiny one should observe their effects
by measuring the Higgs masses and also the Higgs couplings to
ordinary fermions. This MSSM generalisation as pointed out
in\cite{Polonskihardterms} can allow for a lightest Higgs mass in
excess of $150$GeV, say, independently of $\tb$.

\subsection{The Higgs self-couplings}
Some of the expressions below, for the Higgs trilinear couplings,
have been given elsewhere\cite{HaberHempfling,Muhlthesis}. For
completeness we list all the couplings. Introducing

\beqn
\label{gh}
g_h=\frac{2 M_W}{g}= \sqrt{2} v
\eeqn
we have

%%%hhh%%%%
\beqn
\label{hhhc}  g_{hhh}&=& 3 g_h \lambda_{hhh} \nonumber
\\
 \lambda_{hhh}&=&-\frac{e^2}{\stw^2} \sbpa \cta \nonumber \\
&+&
 \lone \; \cb \sa^3-\ltwo\; \ca^3 \sb+
\frac{1}{2}
 (\lthree+\lfour+\lfive)\; \sta \cbpa \nonumber \\
 &+& \lsix \sa^2 (\cbpa +2 \ca \cb) + \lseven \ca^2 (\cbpa -2\sa
 \sb)\\
\label{Hhhc}
g_{Hhh}&=& - 3 g_h \lambda_{Hhh} \nonumber \\
 \lambda_{Hhh}&=&\frac{e^2}{\stw^2} \left(\sbpa \sta -\frac{1}{3}\cbma \right)
 \nonumber \\
&+& \lone \; \cb \sa^2 \ca + \ltwo\; \sb \ca^2 \sa
 -\frac{1}{2}
(\lthree+\lfour+\lfive)\; (\sta \sbpa-\frac{2}{3} \cbma) \nonumber \\
 &+& \lsix \sa (\cb \cta  + \ca \cbpa) - \lseven \ca (\sb \cta  +\sa
 \cbpa)\\
%%%%hHH %%%%%%%
g_{hHH}&=&  3 g_h \lambda_{hHH} \nonumber \\
 \lambda_{hHH}&=&\frac{e^2}{\stw^2} ( \cta \sbpa-\frac{2}{3}\sbma)\nonumber \\
&+& \ca \sa (\lone \ca \cb - \ltwo\ \sa \sb)
-\frac{(\lthree+\lfour+\lfive)}{2}\; ( \sbpa \cta-\frac{\sbma}{3})
\nonumber \\ &+& ( \lsix \ca (\cb \cta -\sa \sbpa) + \lseven \sa
(\cta \sb +\ca \sbpa)\nonumber \\
g_{hAA}&=&  g_h \lambda_{hAA}
\nonumber \\
 \lambda_{hAA}&=&-\frac{e^2}{\stw^2} \sbpa \ctb \nonumber \\
&+& \lone \; \cb \sb^2 \sa- \ltwo\; \sb \cb^2 \ca -
(\lthree+\lfour+\lfive)\; (\sb^3 \ca-\sa \cb^3) + 2 \lambda_5 \sbma \nonumber \\
 &+& \lsix \sb (\ctb \sa + \cb \sbpa) + \lseven \cb (\ctb \ca -\sb
 \sbpa)\\
%%%%hH+H-
g_{hH^+H^-}&=& g_h \lambda_{h H^+ H^-} \nonumber \\
 \lambda_{h H^+ H^-}&=& \lambda_{hAA}-\frac{e^2}{2 s_W^2} \sbma
+ (\lfour-\lfive) \sbma \\
%%%%HHH
g_{HHH}&=& - 3 g_h \lambda_{HHH} \nonumber \\
 \lambda_{HHH}&=&\frac{e^2}{\stw^2}\cta \cbpa \nonumber \\
&+& \ca^3 \cb \lone + \sa^3 \sb  \ltwo
+\frac{(\lthree+\lfour+\lfive)}{2}\; \sta \sbpa
\nonumber \\
&-&\lsix \ca^2 (\sb \ca +3\sa \cb) -\lseven \sa^2 (\cb \sa + 3 \ca \sb)\nonumber \\
%%%%HAA
g_{HAA}&=& - g_h \lambda_{HAA} \nonumber \\
 \lambda_{HAA}&=&
-\frac{e^2}{\stw^2} \cbpa \ctb \nonumber \\
&+&
 \lone \; \cb \sb^2 \ca + \ltwo\; \sb \cb^2 \sa
 +
(\lthree+\lfour+\lfive)\; (\sb^3 \sa + \cb^3 \ca) -2 \lfive \cbma \nonumber \\
 &+& \lsix \sb (\cb \cbpa  + \ca \ctb) - \lseven \cb (\sb \cbpa  +\sa
 \ctb) \\
%%%%HH+H-
g_{HH^+H^-}&=& - g_h \lambda_{H H^+ H^-} \nonumber \\
 \lambda_{H H^+ H^-}&=& \lambda_{HAA}
+\frac{e^2}{2 s_W^2} \cbma
 + (\lfive-\lfour) \cbma
\eeqn

 Written this way the expressions are not very
telling, moreover it is clear that some of the couplings must be
related since after having measured the masses, there remains only
three independent $\lambda$ while there are 8 Higgs trilinear
couplings.

\setcounter{equation}{0}
\section{Expressing the self-couplings in the mass basis}
The writing of the Higgs self-couplings in terms of the
fundamental parameters $\lambda_i$ is not the most judicious. The
reason is that all these parameters, though in different
combinations, already appear in the expression for the Higgs
masses. Therefore the Higgs masses already constrain the Higgs
potential, even if partially. Since the Higgs masses (or at least
the lightest Higgs) will be measured first, the trilinear Higgs
self-couplings being accessed only trough double Higgs production
whose cross section is small, one should ask how much more do we
learn from the measurement of the trilinear couplings once we have
measured the masses. For example, it is quite likely that the
heavy Higgses are too heavy so that one only has access to the
lightest \cpviol even Higgs. In this situation one would have an
extremely precise determination of its mass (either at the LHC or
a linear collider) and also a very precise determination of its
couplings to fermions and (gauge bosons) and hence of the angle
$\alpha$ (or the combination $\alpha -\beta$) at a linear
collider. One should therefore use this information, namely trade
these precisely measured physical quantities for two of the
parameters $\lambda_i$ and then re-express the self-coupling $h/H
\;hh$, in terms of these physical parameters. Of course the choice
of the $\lambda_i$ is not unique, however the parameterisation of
the self-couplings in terms of masses will be more transparent and
would have the advantage of including information on some
previously measured quantities. We therefore propose to use the
physical basis, using as input all the Higgs masses and the mixing
angle $\alpha$\footnote{In another context and in a more
restricted 2 Higgs doublet model, using the ``corrected" masses to
express the self-couplings has been advocated in
\cite{oldselfh-mass,unitarity2hdm}.}. For the latter, beside
$\tb$, one can use the quantities $\sbma$ (and $\cbma$) which can
be directly extracted from the Higgs couplings to fermions or the
vector bosons. For instance a good measurement of the production
cross section of $h$ at \epemt furnishes $\sbma$. The angle
$\beta$ may be measured in some purely non-Higgs processes or if
one has access to some heavy Higgses also through a study of their
couplings to matter \footnote{Of course some of these couplings
receive genuine vertex corrections that are {\em not} encoded in
the correction to the angle $\alpha$, these genuine vertex flavour
dependent corrections should be, whenever possible, subtracted
before one attempts to extract $\sbma$ for example.}. Thus apart
from $M_A$, one can trade $M_h, M_H, M_{H^\pm}, \alpha$ for, for
example, $\lambda_1, \lambda_2, \lambda_3, \lambda_4$ through
\begin{eqnarray}
\label{tradeoffl4}
\lambda_4 & = & (M_A^2+M_W^2-M_{H^{\pm}}^2)/v^2+\lambda_5\\
\label{tradeoffl1} \lambda_1 & = &
\frac{M_h^2s_\alpha^2+M_H^2c_\alpha^2
                -M_Z^2c_\beta^2-M_A^2s_\beta^2}{2v_1^2}
                - \lambda_5t_\beta^2 + 2\lambda_6t_\beta \\
\label{tradeoffl2} \lambda_2 & = &
\frac{M_h^2c_\alpha^2+M_H^2s_\alpha^2
                -M_Z^2s_\beta^2-M_A^2c_\beta^2}{2v_2^2}
                - \lambda_5/t_\beta^2 + 2\lambda_7/t_\beta \\
\label{tradeoffl3} \lambda_3 & = & \frac{(M_H^2-M_h^2)c_\alpha
s_\alpha + (M_A^2+M_Z^2)c_\beta s_\beta}{2v_1v_2}
                - \lambda_4 + \lambda_6/t_\beta + \lambda_7t_\beta.
\end{eqnarray}

 With this choice the different Higgs self
couplings are given in terms of the Higgs masses and the remaining
$\lambda_{5,6,7}$. This choice of basis seems to be the most
natural, since the structures that affect $\lambda_{1-4}$ appear
already at ``tree-level" and thus should be the most affected by
radiative corrections. This is so in the MSSM where the largest
corrections at moderate $\tb$ is in $\lambda_2$, see
\cite{CarenaWagner_Higgs_Approx1} for example. Let us first define
the combinations,
\beqn
\lambda_a&=&\lambda_5 \sinbb
-\frac{\lambda_6+\lambda_7}{2}-\frac{\lambda_6-\lambda_7}{2}\cosbb
\;\;{\rm ( note \; that\;\;} m_{12}^2=-M_A^2 \sb \cb -v^2 \lambda_a {\rm )}\nonumber \\
\lambda_b&=&\lambda_5 \cosbb +\frac{\lambda_6-\lambda_7}{2}\sinbb
\eeqn

After some algebra, where we try as much as possible to express
the $\alpha$ dependence through $\alpha-\beta$, we find

%%%hhh%%%%
\beqn
\label{lhhh}
 \lambda_{hhh}
&=&-\frac{1}{2 v^2} \sbma \; M_h^2 \nonumber \\
 &+&\frac{\cbma^2}{\sb \cb} \left(\sbma \lambda_a + \cbma
 \lambda_b +\frac{(M_A^2-M_h^2)}{2 v^2} (\sinbb \sbma +\cosbb \cbma )\right)
\eeqn

This shows that in the decoupling limit the $hhh$ coupling is
completely determined from the measurement of the Higgs mass,
$M_h$ and the coupling to $WWh/ZZh$, $\sbma$, both of which should
be determined quite precisely from $\epem \ra Zh$ at the LC. The
$\lambda_a$ and more so $\lambda_b$ are totally screened. The
result in the decoupling limit is no surprise as it reproduces
exactly the \sm result. In this regime one essentially have only
one Higgs doublet and if one restricts oneself to dim-4 operators,
one reproduces the \sm exactly.

%%%Hhh%%%%
\beqn
\label{lHhh}
\lambda_{Hhh}&=& \cbma \left\{\frac{4 \lambda_a}{3
\sinbb}
 (1-\frac{3}{2}\cbma^2)+\frac{2 \lambda_b}{\sinbb} \cbma \sbma
 \;+\; \frac{1}{6 v^2}  \left( (3 M_H^2-2 M_h^2) \right. \right.
 \nonumber \\
 &+& \left. \left. 4
 (M_A^2-M_H^2)+\frac{\cbma(\sinbb \cbma-\cosbb \sbma)}{\sb \cb} (2 M_h^2 + M_H^2-3 M_A^2)\right) \right\}
 \nonumber \\
\eeqn

In the decoupling limit

\beqn
 \lambda_{Hhh}= \frac{\cbma}{2 v^2} M_H^2
\eeqn

Since this coupling could most easily be extracted from $H \ra
hh$, $M_H$ would have (together with $\cbma$) already been
measured, thus the new physics effects are again screened.

The remaining couplings involve at least two heavy Higgses an thus
would be more difficult to measure. Nonetheless let us express
them in terms of masses. All the couplings will be written as
\beqn
g_{H_i H_j H_k}=\frac{g_h}{\stb} \lambda'_{H_i H_j H_k}
\eeqn

with
\beqn
 \lambda'_{hAA}&=&2 \sbma\lambda_a+ 2\cbma \lambda_b \;+\;\frac{1}{2 v^2}
\left( \cbma \ctb (2 M_A^2-2 M_h^2) -M_h^2 \stb \sbma \right) \\
 \lambda'_{hHH}
&=& \sbma \left\{ \frac{1}{2 v^2} \left[ \left(
 M_h^2 + 2 (M_H^2-M_A^2) \right) \sta -2 M_A^2 \cbma \sbpa\right]\right.
 \nonumber \\
&+& \left.2 (\lambda_a (1-3 \cbma^2)+3 \cbma \sbma \lambda_b)
\right\} \\
 \lambda'_{hH^+H^-}
&=& -2 \left\{ \frac{1}{2 v^2} \left[ (\ca \cb^3
 -\sa \sb^3) M_h^2 - (M_A^2-M_{H^\pm}^2)\cbpa - M_{H^\pm}^2 \ctb \cbma \right] \right.
 \nonumber \\
& & \left. -(\sbma \lambda_a +\cbma \lambda_b)  \right\} \\
 \lambda'_{HHH}
&=& -6 \left\{ \frac{1}{2 v^2} \left[ \sbpa \left( (M_H^2-M_A^2)
+\cbma^2 M_A^2\right) -\sa \ca \cbma M_H^2 \right] \right.
 \nonumber \\
& & \left. +\sbma^2 (\sbma \lambda_b - \cbma \lambda_a )\right\}
\eeqn

Note once more that these couplings, like other $H$ couplings to
fermions, sfermions and gauge bosons, can be derived from those of
$h$. For example we do verify that $g_{HHH}=g_{hhh}(h \ra H, M_h
\ra M_H, \sbma \ra \cbma, \cbma \ra -\sbma)$.

\beqn
 \lambda'_{HAA}
&=& -2 \left\{ \frac{1}{2 v^2} \left[
 (\sa \cb^3+\ca \sb^3) (M_H^2-M_A^2) + \cbma \cb \sb M_A^2\right] \right.
 \nonumber \\
 & & \left. +\sbma \lambda_b - \cbma \lambda_a \right\} \\
 \lambda'_{H H^+ H^-}
&=& -2 \left\{ \frac{1}{2 v^2} \left[
 \sbpa (M_H^2-M_A^2) +\cbma \cb \sb (2 M_{H^\pm}^2- M_H^2)\right] \right.
 \nonumber \\
 & & \left. +\sbma \lambda_b - \cbma \lambda_a \right\}
\eeqn

In the physical basis and especially by explicitly displaying the
dependence in the mixing angle $\alpha$ through the combination
$\cbma, \sbma$ shows that for couplings involving at least two
heavy Higgses
\begin{itemize}
\item The dependence in the
parameters $\lambda_i$ only appear through the combination
$\lambda_a$ {\em or} $\lambda_b$. Thus one combination is not
accessed in the trilinear couplings
\item In the decoupling limit ($M_h \ll M_A, M_H, M_A$), it is only $\lambda_a$
which is accessible in  the self-couplings involving two heavy
Higgses ($hAA, hHH$ and $h H^+ H^-$), while in the self-couplings
with heavy Higgses only ($H H H, H AA, H H^+ H^-$) only
$\lambda_b$ may be accessed.
\end{itemize}

Note that the heavy Higgs masses, $M_H, M_A, M_{H^\pm}$, enter the
formulae for the couplings only through combinations involving
mass differences or $M_A^2 \cbma$ so that in effect the
self-couplings do not grow with the heavy Higgs mass as occurs in
the \sm.

\subsection{Quartic Higgs couplings}

To access the remaining combination we need to consider the
quartic couplings. We will write the quartic couplings directly in
the physical basis.

Defining
\beqn
\lambda_c=2 \cbb \lambda_b-\lambda_5=\lambda_5 c_{4\beta}+
\frac{\lambda_6-\lambda_7}{2} s_{4\beta}
\eeqn

\beqn
g_{hhhh}&=&-\frac{3 M_h^2}{2 v^2}(1-\cbma^2) \nonumber \\
&-&\frac{3 }{2 v^2 \sb^2 \cb^2}\cbma^2 \left\{ \cbpa(
\sbma\sbb+\cbpa \cbma^2)M_h^2 + \sa^2 \ca^2 M_H^2 - \cbpa^2 M_A^2
\right\}\nonumber
\\ &+& \frac{3 }{ \sb^2 \cb^2}\cbma^2 \left\{ \sbb
\lambda_a+2 \sbb \cbma \sbma \lambda_b +\cbma^2 \lambda_c \right\}
\eeqn

We see again that in the decoupling region one is not sensitive to
any of the extra couplings, as expected since we recover the \sm
result with only the dim-4 operator. Let us now give the formulae
for some of the other quartic couplings to show that some of the
novel couplings are not screened in all of the quartic couplings.
We will only show the dependence in the extra parameters and will
not give the full dependence in terms of masses as otherwise the
formulae may be too lengthy.

\beqn
\lambda_{hhhH}& \rightarrow & \cbma \left\{ \sbma \sbb \lambda_a +
\cbma( 3 \sbb \lambda_b +2 \cbma \sbma \lambda_c -4 \cbma^2 \sbb
\lambda_b )\right\}
\eeqn

Again, all the anomalous couplings are screened.

\beqn
\lambda_{hhHH}& \rightarrow & \sbb \lambda_a + 6 \sbma \cbma
\left( \sbb  \lambda_b + \cbma \sbma \lambda_c -2 \cbma^2  \sbb
\lambda_b \right)
\eeqn

In $hhHH$ only $\lambda_a$ may be accessible.
\beqn
\lambda_{hHHH}& \rightarrow &\sbma \left\{ \sbma \sbb \lambda_b +
\cbma \left(\sbb \lambda_a +2  \sbma^2 \lambda_c - 4 \cbma \sbma
\sbb \lambda_b \right) \right\}
\eeqn

In $hHHH$ only $\lambda_b$ may be accessible.

\beqn
\lambda_{HHHH}& \rightarrow & \sbma^2 \left( \sbma^2 \lambda_c +
\sbb \lambda_a -2 \cbma \sbma \sbb \lambda_b \right)
\eeqn

In $HHHH$ both $\lambda_a$ and $\lambda_c$ may be accessible but
not $\lambda_b$.

\subsection{Using another parameterisation}
The screening property is a general result which does not depend
on which independent parameters we keep beside the physical
masses.  Had we used another set of independent parameters beside
the masses,  the same phenomenon would have occurred and only two
independent combinations out of the three parameters would enter
the expression of the trilinear couplings. Indeed with
$\lambda_3,\lambda_5,\lambda_6$ as extra parameters, the role of
$\lambda_{a,b,c}$ is played by $\lambda_{a,b,c}'$, such that

\beqn
\lambda_a \ra \lambda_a'&=&-\sb \cb (\lambda_3-\lambda_5) \nonumber \\
\lambda_b \ra \lambda_b'&=&- \left(
\frac{\lambda_3+\lambda_5}{2}+\ctb \frac{\lambda_3-\lambda_5}{2}
\right) +\frac{\cb}{\sb} \lambda_6 \nonumber \\
\lambda_c \ra \lambda_c'&=& 2 \lambda_b' \cbb -\lambda_5
\eeqn

%These combinations can be derived directly from the set
%$\lambda_3,\lambda_5,\lambda_7$ through Eq.~first in paragraph 3.

For instance,
\beqn
 \lambda_{hhh}&=&-\frac{1}{2 v^2} \sbma \; M_h^2
 \; + \;\cbma^2 \frac{1}{\cb \sb}
 \left( \sbma \lambda_a'+ \cbma \lambda_b'\right) \nonumber
\\ &+&
\frac{\cbma^2}{2 v^2 \cb \sb^2}\left\{\sbma \cb \sb^2
(M_A^2-M_H^2-M_h^2-M_Z^2 c_{2W}+2M_{H^\pm}^2) \right. \nonumber
\\ && \;\;-\cbma  \left[ \sb \left( \sb^2 (M_A^2-M_H^2)+\cb^2 (2M_H^2+M_Z^2
c_{2W}-2M_{H^\pm}^2 -M_h^2) \right) \right. \nonumber
\\ & & \;\;\;\;
\left. \left. +\cbma (M_H^2-M_h^2) \left( \sbma \cb (1-4 \sb^2)
-\cbma \sb (3-4\sb^2)\right) \right] \right\}
\eeqn

Compared to the previous parameterisation, this looks rather more
complicated as it involves the charged Higgs mass as well as the
heavy $H$ beside the pseudoscalar Higgs mass. Nonetheless all
these masses are screened.

\setcounter{equation}{0}
\section{Dim-4 operators and independent parameters}
To easily understand our finding about the screening and the
number of independent parameters in the trilinear and quadrilinear
couplings, note that the trilinear  and quadrilinear couplings
originate from the quartic terms $\lambda_i$ only,
Eq.~\ref{potential4}, whereas the mass terms get an additional
contribution from the bi-linear terms $m_1, m_2, m_{12}$ in
Eq.~\ref{potential4}. Take for instance the case of the neutral
couplings. The quartic self-couplings emerge as combinations of
$5$ independent terms in the original fields (before
diagonalisation) of the form
\beqn
h_1^4 \;,\; h_2^4 \;,\; h_1^3 h_2 \;,\; h_1^2 h_2^2 \;,\; h_1
h_2^3
\eeqn
In terms of the physical scalar fields, $h,H$:
\beqn
h_1=-h s_\alpha + H c_\alpha \;\;\;\; h_2=h c_\alpha + H s_\alpha
\eeqn

which we more judiciously write as
\beqn
h_1&=&\cb (\sbma h+\cbma H) +\sb (H\sbma -h\cbma)=\cb h_-+\sb h_+
\nonumber \\
h_2&=&\sb h_--\cb h_+
\eeqn

However keep in mind that  $h_1$ and $h_2$ in Eq.~\ref{doublets}
always appear in the form

\beqn
v_1+\frac{h_1}{\sqrt{2}}=v_1 \left( (1+\frac{h_-}{v'}) + \tb
\frac{h_+}{v'} \right) \nonumber
\\
v_2+\frac{h_2}{\sqrt{2}}=v_2 \left( (1+\frac{h_-}{v'}) - \tb^{-1}
\frac{h_+}{v'} \right)
\eeqn
with $v'=\sqrt{2} v$($=g_h$ Eq.~\ref{gh}), which helps write the
doublets, Eq.~\ref{doublets}, as

\beqn
\label{doubletclever} H_1=v_1 \left(\begin{array}{c} 1 + ((h_- - i
G^0)+\tb (h_+ + i A))/v' \\ (-G^- +\tb H^-)/v
\end{array} \right) \nonumber \\ H_2=v_2 \left(\begin{array}{c}
(G^+ +\tb^{-1}H^+)/v \\ 1+((h_- + i G^0)-\tb^{-1} (h_+ - i A))/v'
\end{array}\right)
\eeqn

where $G^{\pm,0}$ stand for the Goldstone bosons.

 Therefore in effect the quartic terms originate from a
combination of the form

\beqn
(1+\frac{h_-}{v'})^2 \left( q_1 (1+\frac{h_-}{v'})^2 + q_2
(1+\frac{h_-}{v'}) \frac{h_+}{v'} +q_3 \frac{h_+^2}{v'^2} \right)
+ q_4 (1+ \frac{h_-}{v'}) \frac{h_+^3}{v'^3} + q_5
\frac{h_+^4}{v'^4}
\eeqn

While the bi-linear terms are of the form

\beqn
b_1 (1+\frac{h_-}{v'})^2 \;+\; b_2 (1+\frac{h_-}{v'})
\frac{h_+}{v'} \; +\; b_3 \frac{h_+^2}{v'^2}
\eeqn

Imposing that \underline{no tadpole} remains (no linear term in
$h_+, h_-$) means that ($b_1,q_1$) and ($b_2 ,q_2$) must combine
such that one has

\beqn
\label{crunch} 4 q_1 \frac{h_-^2}{v'^2}
(1+\frac{h_-}{v'}+\frac{h_-^2}{4 v'^2}) + 2 q_2
\frac{h_+h_-}{v'^2} (1+\frac{3 h_-}{2 v'}+\frac{h_-^2}{2 v'^2})
\nonumber
\\
+ \frac{h_+^2}{v'^2} (b_3 + q_3 (1+\frac{h_-}{v'})^2) + q_4
(1+\frac{h_-}{v'}) \frac{h_+^3}{v'^3} + q_5 \frac{h_+^4}{v'^4}
\eeqn

 Since the coefficients of $h_-^2 (q_1)$ and $h_- h_+
(q_2)$ are expressed in terms of masses so do those of
$h_-^3,h_-^4$ ($q_1$) as well as $h_+ h_-^2,h_+ h_-^3$ ($q_2$).
This does not apply to the trilinear and quadri-linear terms
involving $h_+^2$ and higher order in $h_+$. Thus for the
trilinear terms there are only the two structures $h_+^2 h_-$
(with coefficient $q_3$) and $h_+^3$ (with coefficient $q_4$) that
can not be expressed solely in terms of the physical masses. The
last parameter $q_5$ only appears in the quartic couplings in the
form $h_+^4$. To obtain $q_1$ and $q_2$ is a straightforward
matter. One only has to rewrite the \cpviol-even Higgs masses in
terms of $h_\pm$:
\beqn
&& M_h^2 h^2 + M_H^2 H^2 \ra \nonumber \\
&& h_-^2 (M_h^2 \sbma^2 + M_H^2 \cbma^2 ) \; +\; 2 h_+ h_- \sbma
\cbma (M_H^2-M_h^2) + h_+^2 (M_H^2 \sbma^2 + M_h^2 \cbma^2 )
\nonumber \\
\eeqn

$q_3$, $q_4$ and $q_5$ are directly related to the parameters
$\lambda_a, \lambda_b, \lambda_c$. The construct of
Eq.~\ref{crunch} shows that in fact once we get the quartic
couplings one also derives the coefficients of the various
trilinear couplings. For example, take the quartic couplings as
they appear in the original potential in terms of the fields $h_1$
and $h_2$,
\beqn
Q_h=\frac{\lambda_1}{2} h_1^4 + \frac{\lambda_2}{2} h_2^4+
(\lambda_3+\lambda_4+\lambda_5) h_1^2 h_2^2 -2 \lambda_6 h_1^3 h_2
- 2 \lambda_7 h_2^3 h_1
\eeqn

When expressed in terms of $h_-$ and $h_+$ and after moving to the
mass basis with for example $\lambda_{5,6,7}$ as extra parameters
we immediately get the following dependence of the various quartic
and trilinear terms

\beqn
\label{compact} & & (h_+^2+ A^2+ 2H^+ H^-)\left[ 2 (\lambda_a h_-
- \lambda_b h_+) +\frac{1}{v'} \left( \lambda_a (h_-^2+G^{0^2}+2
G^+G^-) \nonumber
 \right. \right. \\
& & \left. \left.  -2\lambda_b (h_- h_+-AG^0-(H^+G^-+H^-G^+))
+\tilde{\lambda}_c (h_+^2+ A^2+ 2H^+ H^-) \right) \right]
\eeqn

with \footnote{The reason $\lambda_c$ appears instead of
$\tilde{\lambda}_c$
 depends on how we organise the decoupling and is due to the
 rewriting of the terms in $\sbma^2$ as $1-\cbma^2$. }
\beqn
\tilde{\lambda}_c=\frac{c_{2\beta}^2}{s_{2\beta}}\lambda_5
-\frac{\lambda_6+\lambda_7}{2}
+\frac{\lambda_6-\lambda_7}{2}c_{2\beta} \;{\rm and} \;
\tilde{\lambda}_c-\lambda_a=\frac{1}{s_{2\beta}} \lambda_c
\eeqn

In fact with this parameterisation one has that $q_3=\lambda_a$,
 $q_4=\lambda_b$ and $q_5=\tilde{\lambda}_{c}$. One can move to another
 parameterisation, {\it i.e} choosing a different set of extra parameters,
 through Eqs.~\ref{tradeoffl4}-\ref{tradeoffl3}.
We see that by {\em completing} the $h_-,h_+$ dependence of
$H_{1,2}$ we even get the full trilinear and quadri-linear
couplings involving the pseudo-scalar, charged and Goldstone Boson
couplings.  The completion is obtained by identifying the
different $\tb$ dependencies (namely $1, \tb, \tb^2$) in the
modulus of $H_1$ for instance. It is not hard to see that by
re-expressing $h_-$ and $h_+$ in terms of $h$ and $H$,  we recover
all our results. Moreover this writing immediately shows that
trilinear scalar couplings involving Goldstone Bosons can all be
expressed most simply in terms of the physical Higgs masses only.
The requirement for the absence of tadpoles is a crucial one and
explains most of our findings when restricting ourselves to dim-4
operators.

\setcounter{equation}{0}
\section{Radiative Corrections}
Our results can also be exploited for easily expressing the
radiative corrections to the trilinear (and for that matter
quadrilinear) couplings of the SUSY Higgses and explain some of
the properties pointed out in the literature. Three-point one-loop
radiative corrections for the neutral Higgs system in the MSSM
have been calculated \cite{RCmhtrilinear} within the effective
potential approximation\footnote{See also \cite{Panditatrilinear}
where expressions are given for some couplings assuming equal soft
masses for the stop masses. Note however that there is a misprint
in Eq.~2.4 of \cite{Panditatrilinear} where in the last term of
that equation one should read $A (A+\mu \cot \beta)$ instead of
$\mu (A+\mu \cot \beta)$.}. The diagrammatic one-loop radiative
corrections to both the trilinear, $\lambda_{hhh}$ and
quadrilinear $\lambda_{hhhh}$ lightest Higgs self couplings have
been revisited recently in\cite{Holliksusyselfcouplings}. For the
case of no mixing in the stop sector it is shown
analytically\cite{Holliksusyselfcouplings} that the bulk of the
corrections in the couplings are absorbed by using the corrected
Higgs mass while the same is demonstrated numerically for the case
of large mixing. One-loop radiative corrections for
$\lambda_{hAA}$ are also considered in \cite{brignole1,brignole2},
for $\lambda_{HAA}$ in \cite{brignole1}  and for $\lambda_{Hhh}$
in \cite{brignole3}. Here also the corrections are found to be
large before re-expressing the results in terms of the corrected
masses.

Ref.~\cite{CarenaWagner_Higgs_Approx1} gives analytical
approximations (including 2-loop leading-log corrections) for the
effective quartic couplings $\lambda_i$\footnote{Compared to our
notations we should make $\lambda_{6,7} \ra -\lambda_{6,7}$ in the
expressions of \cite{CarenaWagner_Higgs_Approx1}. Moreover our
sign convention for $\mu$ is the opposite of
\cite{CarenaWagner_Higgs_Approx1} but the same as in
\cite{Panditatrilinear}. See \cite{nous_RggstopHiggs_lhc} for a
full definition of our conventions.}, using a
renormalisation-group improved leading-log approximation. We can
adapt their formulae to the one-loop case with  large stop masses
so that we can compare with the direct calculation of the vertices
performed in \cite{RCmhtrilinear,Panditatrilinear}. For instance,
we find the leading one-loop contributions in the limit where the
SUSY breaking term, $M_S^2$, defined below, is (much) larger than
the top mass, (which is consistent with our approach of keeping
only the dimension-4 operators)

\beqn
\label{rcdhhh}
 \Delta \lambda_{hhh}&=&- \frac{g^4 m_t^4}{64 \pi^2 M_W^4}
\frac{\ca^3}{\sb^3} \left\{ 6 \log (M_S^2/m_t^2) \; +3 \frac{f_t
(c_t+f_t)}{M_S^2} - \frac{c_t f_t^3}{2 M_S^4}  \right\} \nonumber
\\ \Delta \lambda_{Hhh}&=& \frac{g^4 m_t^4}{64 \pi^2 M_W^4}
\frac{\sa \ca^2}{\sb^3} \left\{ 6 \log (M_S^2/m_t^2) \; +
\frac{c_t (e_t +2 f_t)+ f_t (f_t+2 e_t)}{M_S^2} - \frac{c_t e_t
f_t^2}{2 M_S^4} \right\} \nonumber \\ c_t&=& A_t+\mu/\tb \;\;\;
e_t=A_t+\mu/\ta \;\;\; f_t=A_t-\mu \ta \;\;\; m_{\sto ,
\stt}^2=M_S^2\pm m_t c_t
\eeqn

 These shifts correct the
tree-level expression of Eq.~\ref{hhhc} and Eq.~\ref{Hhhc}
respectively. In the limit $\mu \ra 0$, all the $\lambda_i$ in the
MSSM vanish but $\lambda_2$:
\beqn
\lambda_2^{\sto,\stt}&=&\frac{3}{32\pi^2} \frac{g^4 m_t^4}{\sb^4
M_W^4} \left\{ \log(M_S^2/m_t^2) + \frac{A_t^2}{M_S^2}(1-A_t^2/12
M_S^4)\right\}\nonumber
\\ & \sim & .15 \; {\rm for} \; A_t=M_S=1TeV \;\;{\rm and} \;\;\tb=10
\eeqn

while in the same approximation as Eq.~\ref{rcdhhh}
\beqn
M_h^2&=&M_Z^2\sbpa^2 +M_A^2 \cbma^2 + \Delta M_h^2 \nonumber \\
\Delta M_h^2&=& \frac{3}{8 \pi^2} \frac{g^2 m_t^4}{M_W^2}
\left(\log (M_S^2/m_t^2) \; +\frac{f_t c_t}{M_S^2} - \frac{c_t^2
f_t^2}{12 M_S^4}  \right) \left(
1+\frac{\cbma}{\sb^2}(s_{2\beta}\sbma+ c_{2\beta}\cbma)\right)
\nonumber \\
\eeqn

\noindent so that one recovers the decoupling property and the
fact that the bulk of the radiative corrections are reabsorbed by
using the corrected Higgs mass,
\beqn
\Delta \lambda_{ hhh}=-\frac{\Delta M_h^2}{2 v^2} \left(\sbma
+\frac{\cbma}{\tb} \right) \;+\; \frac{3  m_t^4}{16 \pi^2 v^4}
\frac{\mu f_t}{M_S^2} \left(1-\frac{f_t c_t}{6 M_S^2}\right)
\frac{\ca^2}{\sb^4} \cbma
\eeqn

In a phenomenological analysis of the extraction of the Higgs
self-couplings, one could add the contribution of the stop-sbottom
at the two-loop level through effective couplings $\lambda_i$ from
the renormalisation group improved results of
\cite{CarenaWagner_Higgs_Approx1} to which one could include new
physics contributions to the $\lambda_i$.

Note that contrary to what we have presented in the previous
sections, we have shown the ``corrections" to the Higgs
self-couplings due to radiative corrections (or presence of
$\lambda$ terms) as shifts compared to the tree-level MSSM. We
have done so in order to compare with the existing
literature\cite{RCmhtrilinear,Panditatrilinear,Holliksusyselfcouplings}
that takes into account effects at one-loop only. Although this
shows that the bulk of the corrections is absorbed in terms of the
Higgs mass, the notion of shifts here is somehow misleading
especially that some of the correction is contained in the
``corrected" mixing angle $\alpha$.

\setcounter{equation}{0}
\section{Phenomenology and reconstruction of the Higgs potential at future colliders}
\begin{figure*}[thbp]
\begin{center}
\mbox{\includegraphics[height=8cm,width=0.5\textwidth]{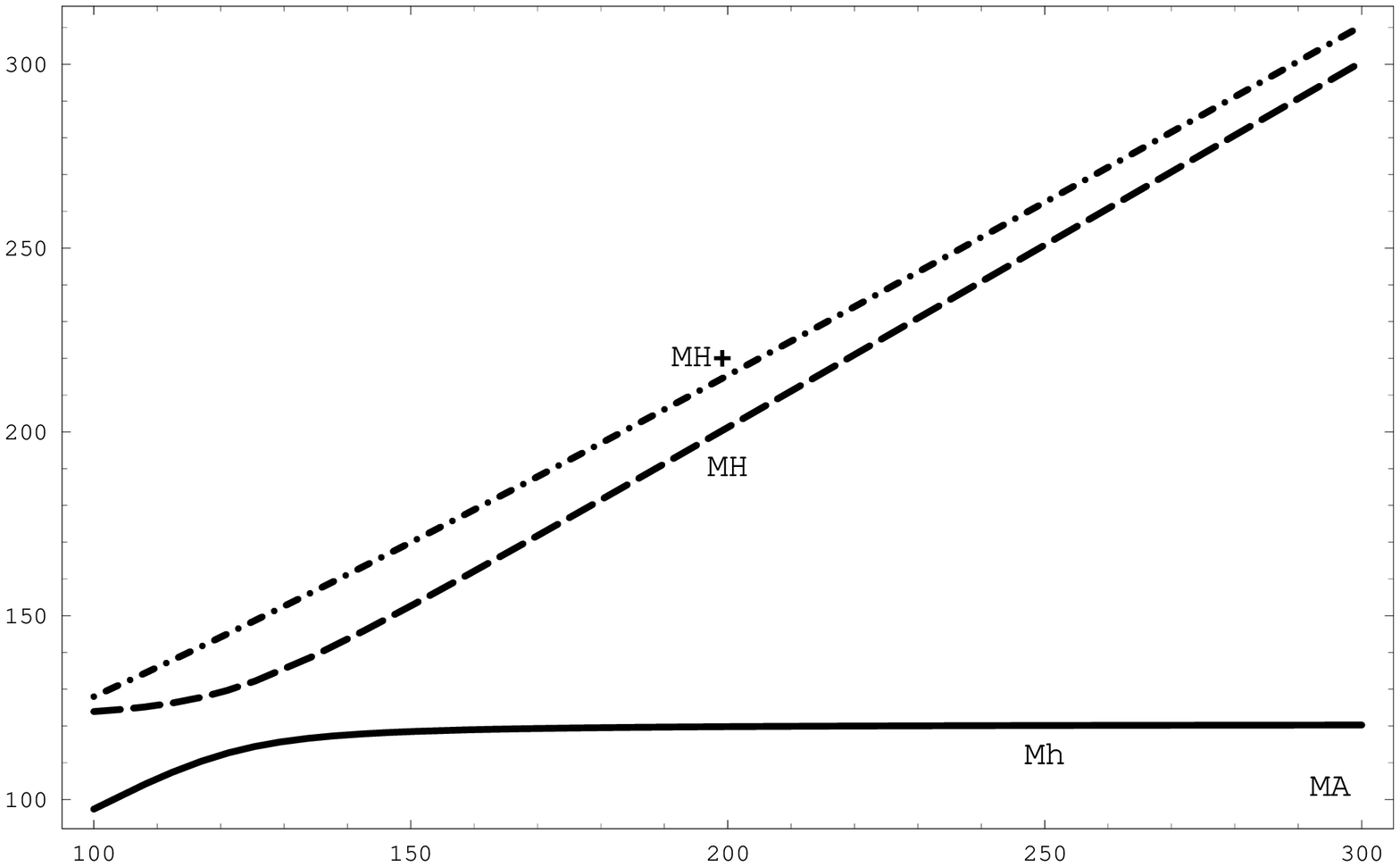}
\includegraphics[height=8cm,width=0.5\textwidth]{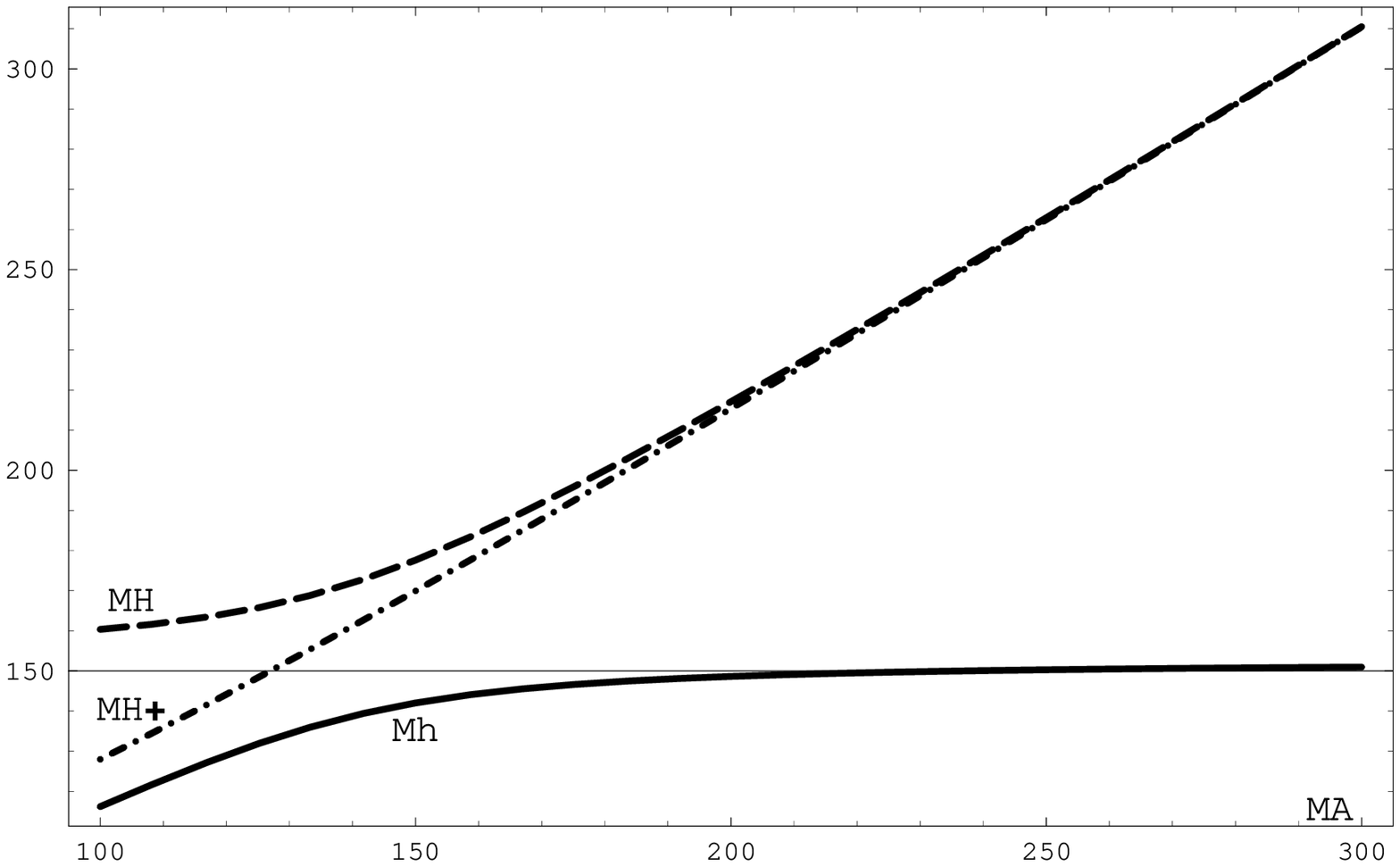}}
\caption{\label{figsmasses}{\em  Higgs mass spectrum without
``hard" terms but with  $A_t=1000$GeV, $M_S=800$GeV, $\mu=-300$GeV
and $\tgb=10$ (left) and with the inclusion of additional terms
with $\lambda_{1-5}^{\rm new}=-\lambda_{6,7}^{\rm new}=.1$(right).
All masses are in GeV. }}
\end{center}
\end{figure*}

As we have seen, the measurement of the entire set of the dim-4
operators which is necessary to reconstruct the Higgs potential in
SUSY (and 2HDM) requires that one crosses the thresholds for the
production of 3 Higgs bosons, which is not an easy task especially
that the cross sections will get tinier and tinier as the Higgs
multiplicity increases. As we have seen also, a precise
measurement of the Higgs masses and their couplings to ordinary
matter is an important ingredient in the reconstruction of this
potential. The LHC can thus give a first hint on the parameters
$\lambda_i$. For instance imagine that the LHC discovers some SUSY
particles and identifies them as such but that one discovers also
that the lightest Higgs has a mass in excess of $150$GeV. This
would point to a scalar potential with ``hard" $\lambda$ terms. We
could probably even set a rough bound on their possible values. A
LC with enough energy to produce some of the Higgses and good
luminosity to probe their couplings would constitute a nice
complementary machine though. Although double Higgs production at
the
LHC\cite{trilinearHiggsMSSMlhc,Lafayehhh,hpluslhc,hcouplingsLHC,Muhlthesis}
may not be so negligible, extracting the trilinear Higgs
self-couplings will prove a challenge\cite{Lafayehhh}. Therefore
for the rest of this session we will only briefly outline what
might be measured from the self-couplings of the Higgs at
different stages of the LC. However, before doing so, let us
illustrate what the mass measurements alone can bring and how the
spectrum can be drastically affected by different forms of the
potential. As an illustration we stick to $\tgb=10$ and consider
the situation where the $\lambda$ receive corrections from the
stop sector with the following parameters $A_t=1000$GeV,
$M_S=800$GeV,$\mu=-300$GeV. We will compare the situation where no
``hard" terms are added with a situation with
$\lambda_{1-5}=-\lambda_{6,7}=.1$, {\it i.e.} of the order of
$\lambda^{\sto \stt}_2$. The mass spectrum of the Higgs system for
this choice of parameters is shown in Fig.~\ref{figsmasses}
\begin{figure*}[bhtp]
\begin{center}
\includegraphics[height=8cm,width=.9\textwidth]{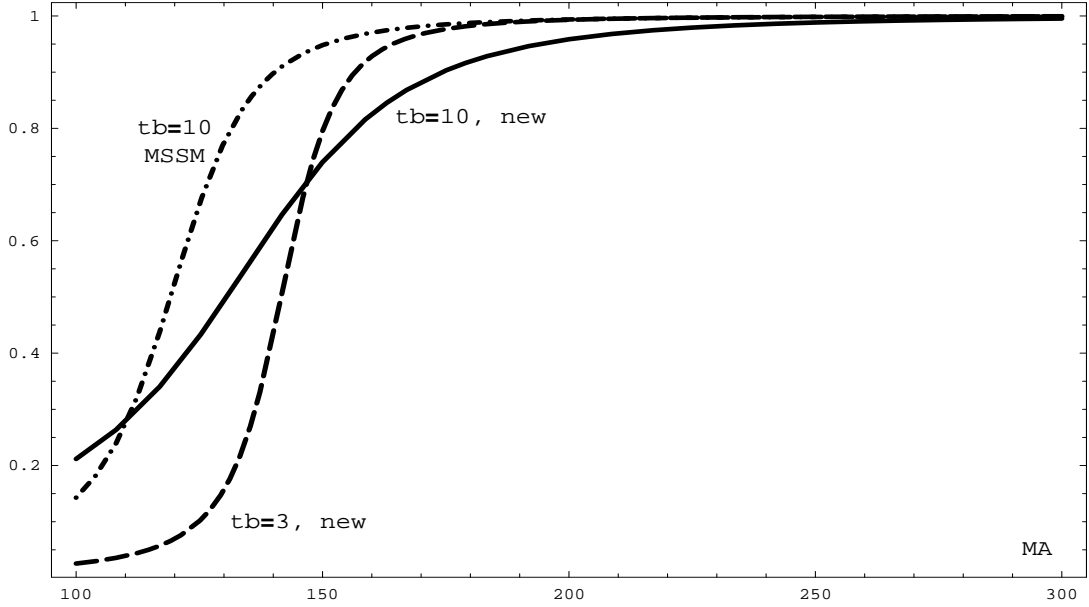}
\caption{\label{figangles}{\em  $\sbma^2$ as a function of $M_A$.
MSSM refers to $\lambda_i=0$ with $A_t=1000$GeV,
$M_S=800$GeV,$\mu=-300$GeV while ``new" has $\lambda_{1-5}^{\rm
new}=-\lambda_{6,7}^{\rm new}=.1$. tb stands for $\tgb$.}}
\end{center}
\end{figure*}

One striking feature is that $M_h$ can be substantially heavier
than what it is in the usual MSSM, while the mass ordering between
$M_{H^{\pm}}$ and $M_H$ is certainly another distinguishing
feature for this particular choice of parameters.

The rate of $h$ production at \epemt, weighted by $\sbma^2$, can
also provide a helpful hint and additional constraint. However,
decoupling although slightly delayed by the presence of the new
$\lambda_i$, occurs rather fast in this variable as shown in
Fig.~\ref{figangles}. Having measured $\tgb$ greatly helps as the
figure illustrates.

A full analysis from the measurements of the masses and the
couplings to fermions and vector bosons is left to a forthcoming
detailed analysis\cite{wehhhexp}.

\begin{figure*}[hbtp]
\begin{center}
\includegraphics[height=9cm,width=1.\textwidth]{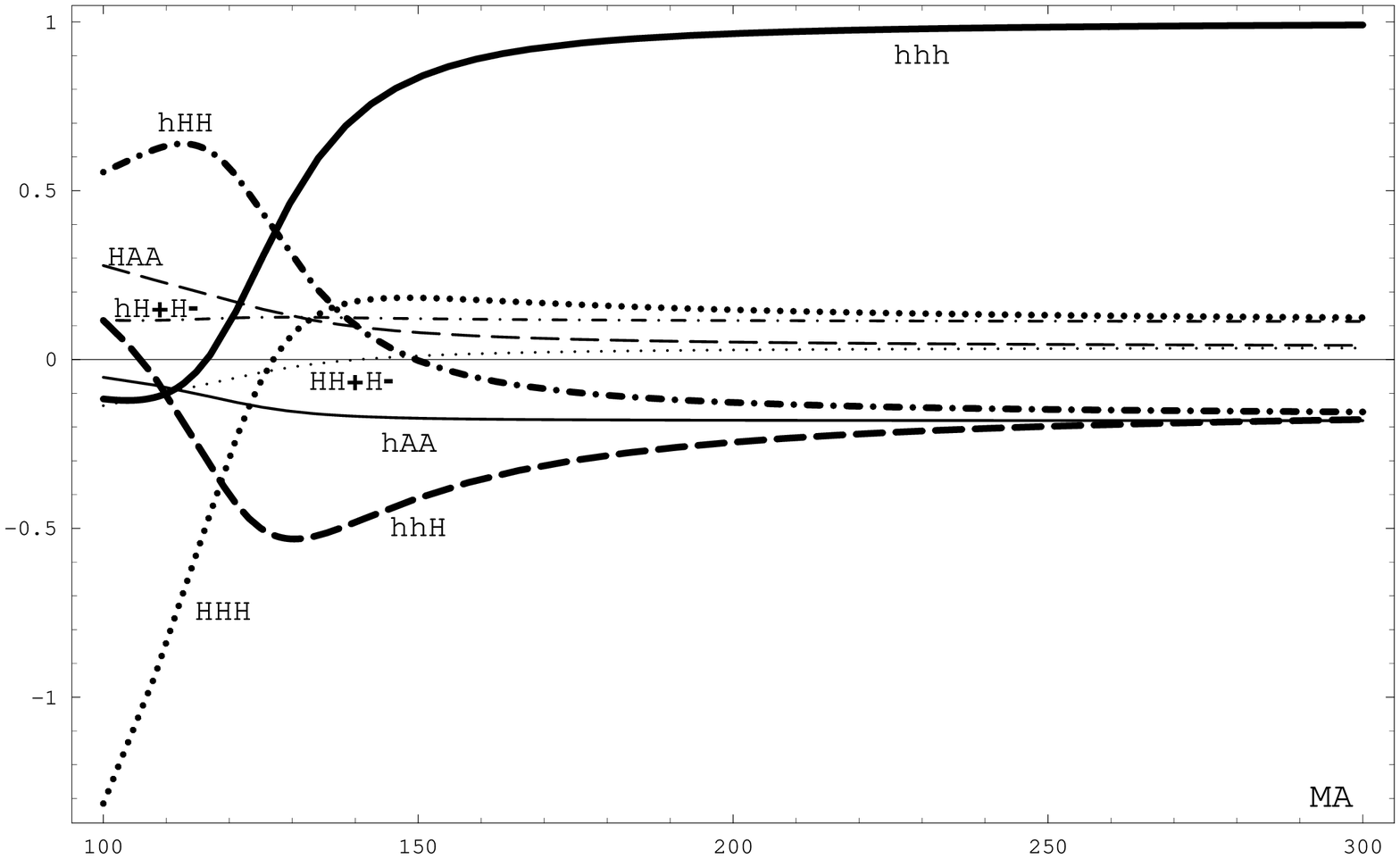}
\includegraphics[height=9cm,width=1.\textwidth]{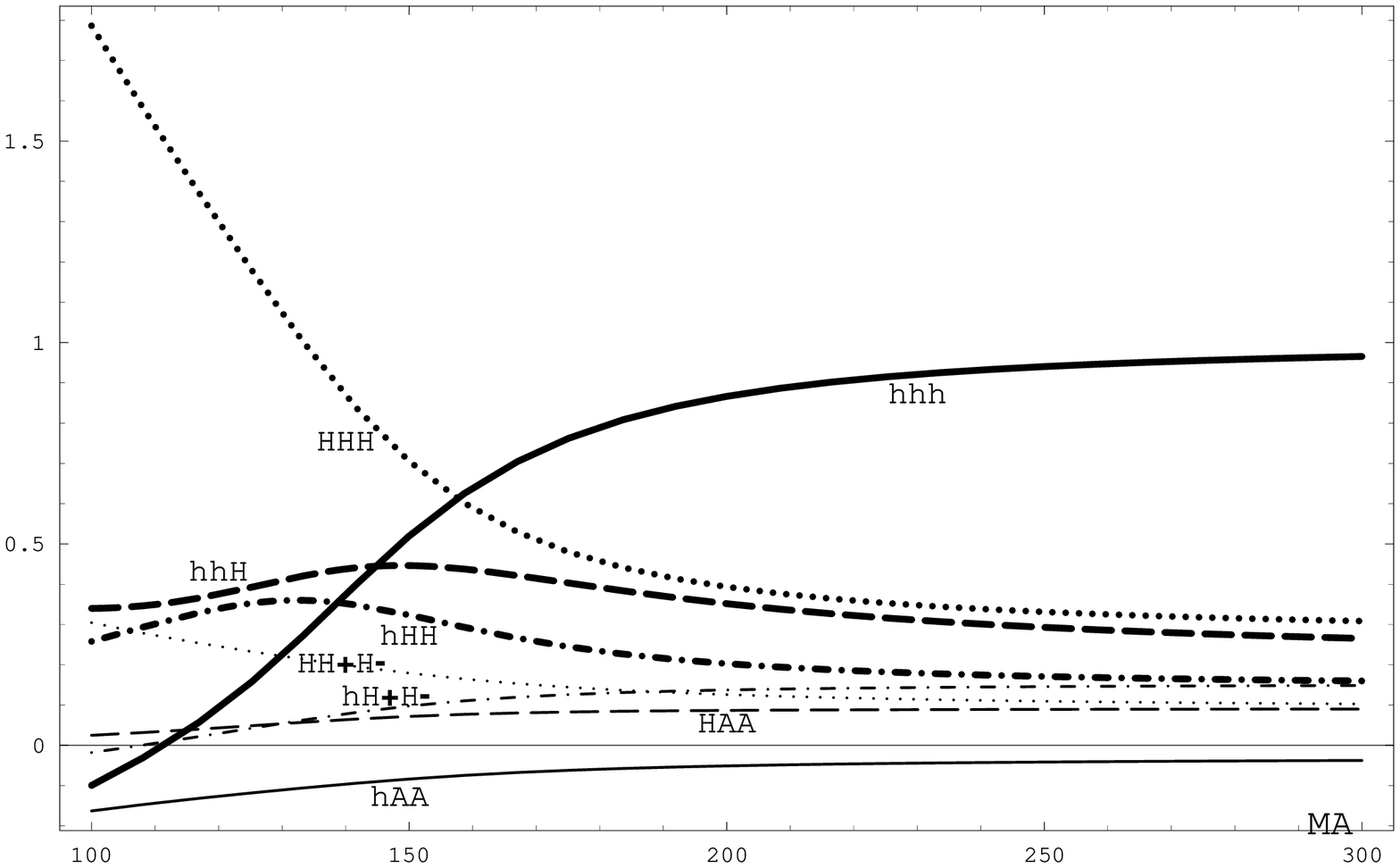}
\caption{\label{fighhh}{\em  $g_{H^i H^j H^k}/\tilde{g}_{hhh}^{\rm
SM}$ with $\tilde{g}_{hhh}^{\rm SM}=-3M_h^2/2v^2$. The SM Higgs
mass is identified with $M_h$ (and thus varies with $M_A$). The
curve in the first panel is the usual MSSM where the one in the
second panel is defined with the same parameters as in
Fig.~\ref{figsmasses}.}}
\end{center}
\end{figure*}

As for the Higgs self-couplings the presence of $\lambda_i$ can
have a drastic effect as shown in Fig.~\ref{fighhh} especially for
small $M_A$, see in particular the swing in $g_{HHH}$. In this
region one expects though that all Higgs masses would have been
measured and thus a good constraint on the parameter space will
have been provided by the masses. As soon as we enter the
decoupling region, the largest coupling is $g_{hhh}$ which reaches
its SM value. The other couplings remain unfortunately rather
small although some have values larger than their corresponding
\sm value. However as we have seen the bulk of these deviations is
due to the rather large deviations in the Higgs masses. In this
respect let us note that  Fig.~\ref{fighhh} seems to indicate that
the $HHH$ coupling can get rather large for small $M_A$. However
observe that we have plotted a reduced coupling in units of the SM
coupling $\tilde{g}_{hhh}^{\rm SM}=-3M_h^2/2v^2$. The reason the
reduced coupling attains a value larger than 1 is due to the
larger mass of H and that we are in a region of non decoupling,
see Fig.~\ref{figangles}. In this region $H$ is more standard-like
than $h$, as far as its couplings to gauge bosons are concerned.
Had we used $\hat{g}_{hhh}^{\rm SM}=-3M_H^2/2v^2$ as a unit, the
reduced coupling would be below 1.

Let us now review briefly how an \epemt machine working at
successive thresholds for Higgs production can attempt to unravel
the Higgs potential.

\subsection{Stage 1}
Imagine a situation where no heavy Higgs has been produced at a
first stage of a linear collider at $500$GeV or the LHC, we would
then be in the decoupling limit. The only trilinear couplings
which may be accessed are $hhh$ and $Hhh$ through $\epem \ra Zhh$
(fusion channels are not efficient at these energies and Higgs
masses). However there is no sensitivity to $Hhh$. Indeed the
amplitude for $\epem \ra Zhh$, in the unitary gauge, can be
written as

\beqn
\label{eezhh}
 {{\cal M}}_{Zhh}=a_h \lambda_{hhh} \sbma  \;+\; a_H
\lambda_{Hhh} \cbma  + R_a
\eeqn
where $R$ stands for  other  contributions not containing the
trilinear Higgs couplings.  We have seen that $\lambda_{Hhh}$ is
screened by a factor $\cbma$ (Eq.~\ref{lHhh}), it is further
screened by another such factor when we consider its contribution
to this cross section.

At this stage the best would be to reconstruct as precisely as
possible $M_h$ and the couplings of $h$ to fermions and the vector
bosons. This will help give a bound on the $\lambda_i$. If one
makes some model dependent assumptions on the $\lambda_i$(imposing
some discrete or global symmetries), this can be used to extract
some information on $M_A$. If an independent measurement of $\tgb$
is missing at the time of the measurements of the Higgs
properties, this will complicate the analyses.
\\ \noindent If, on the
other hand,  the mixing angle is such that $\cbma$ is not too
small and that $M_A$ is not too large, $\epem \ra Z H$ may be
accessible. Then the coupling $\lambda_{Hhh}$ could be reached
directly through $H \ra hh$. This may still turn out not to be too
helpful, since we have seen that the $\lambda_i$ are still
screened in this coupling, even though the screening in this
situation could be mild. Moreover $H$ decays into other particles
($t\bar t$ or $b\bar b$,..) and superparticles (charginos and
neutralinos) may still be dominant so that $Br(H\ra hh)$ will be
poorly determined. Let us remark at this point that most of the
nice analyses of the SUSY Higgs
self-couplings\cite{Teslaphysics,hhhbattagliaboos} that have been
performed were done solely in the context of the minimal
supersymmetric model, with no additional ``hard" terms in the
potential, and have relied heavily on the extremely good precision
of the measurement of the dominant branching ratio into $b\bar b$.
In case $h$ is heavier than $150$GeV these analyses need to be
extended.

\subsection{Stage 2}
For a machine with higher energies where $H$ and $A$ and thus most
probably  $H^\pm$ have been discovered, the first thresholds for
double Higgs production (after that of $Zhh$) may be

\beqn
&&\epem \ra ZhH, \nu_e \bar{\nu}_e Hh  \;{\rm with}\;\; {{\cal
M}}_{Hh}=b_h \lambda_{Hhh}
\sbma \;+\; b_H \lambda_{HHh} \cbma  + R_b \nonumber \\
&&\epem \ra hh A  \;{\rm with}\;\; {{\cal M}}_{Ahh}=c_h
\lambda_{hhh} \cbma \;+\; c_H \lambda_{Hhh} \sbma + c_A
\lambda_{hAA} \cbma + R_c
\eeqn

Again, unfortunately these two reaction will not be very sensitive
to deviations in the trilinear couplings if one takes into account
the screening effect in $\lambda_{Hhh}$. Fusion processes could
also be exploited at this stage and the next, but they exhibit a
similar behaviour to the annihilation processes as far the
extraction of the parameters is concerned.

%To add?
% the width of $H$ into a pair of light Higgs is
%\beqn
%\Gamma(H \ra hh)=\frac{\sqrt{2} G_F M_Z^4 \beta_h}{32 \pi M_H}
%\lambda^2_{Hhh}
%\eeqn

\subsection{Stage 3}
With higher energies one produces two heavy Higgses in association
with a light Higgs or a Z.

\beqn
&&\epem \ra ZHH, \nu_e \bar{\nu}_e HH \;{\rm with}\;\; {{\cal
M}}_{HH}=d_h \lambda_{HHh}
\sbma \;+\; d_H \lambda_{HHH} \cbma  + R_d \nonumber \\
&&\epem \ra Z A A, \nu_e \bar{\nu}_e A A  \;{\rm with}\;\; {{\cal
M}}_{ZAA}=e_h \lambda_{hAA} \sbma \;+\; e_H \lambda_{HAA} \sbma  +
R_e \nonumber
\\
&&\epem \ra h H A \;{\rm with}\;\; {{\cal M}}_{AHh}= (f_h
\lambda_{HHh}
 \;+\; f_H \lambda_{hAA}) \sbma + (f'_h
\lambda_{Hhh}  + f'_H \lambda_{HAA})\cbma + R_f \nonumber \\
&&\epem \ra H^+ H^- h  \;{\rm with}\;\;  {{\cal M}}_{H^+ H^-
h}=g_h
\lambda_{H^+H^-h} +R_g \nonumber \\
&&\epem \ra Z H^+H^-, \nu_e \bar{\nu}_e H^+H^- \;{\rm with}\;\;
{{\cal M}}_{X H^+H^-}=h_h \lambda_{h H^+ H^-} \sbma \;+\; h_H
\lambda_{H^+H^-} \cbma + R_h\nonumber \\
\eeqn

As can be seen all of these reactions will be used to determine
$\lambda_a$ ($\lambda_b$ will still be screened). Let us give some
idea about the order of magnitude of the cross sections to show
that things can get really tough. As a reference take all extra
contributions to the $\lambda_i$ to be vanishing with SUSY
parameters as those considered in the introduction  of this
section: $A_t=1000$GeV, $M_S=800$GeV, $\mu=-300$GeV and $\tgb=10$
and take $M_A=300$GeV. The third stage could be taken as
$\sqrt{s}=1.2$TeV. We find that $ZHH$ and $ZAA$ are about $2.3\;
10^{-2}$fb, while the other processes listed in this stage are 2
orders of magnitude below. Before taking into account signatures
and efficiencies this can amount to about only $25$ events a year
based on a luminosity of $1$ab$^{-1}$.

\subsection{Stage 4}
At even  higher energies, production of three heavy Higgs bosons
could in principle allow to determine $\lambda_b$. The processes
at our disposal will be

\beqn
\epem \ra AAA & & \;{\rm with}\;\;  {{\cal M}}_{AAA}=i_h
\lambda_{hAA}
\cbma \;+\; i_H \lambda_{HAA} \sbma  \nonumber \\
\epem \ra H H A & & \;{\rm with}\;\;  {{\cal M}}_{H H A}=j_h
\lambda_{hHH} \cbma \;+\; j_H \lambda_{HHH} \sbma +j_A
\lambda_{HAA} \sbma  + R_j \nonumber
\\
\epem \ra H^+ H^- H & & \;{\rm with}\;\; {{\cal M}}_{H^+ H^-
H}=k_H
\lambda_{H^+H^-H} + R_k \nonumber \\
\epem \ra H^+ H^- A & & \;{\rm with}\;\;  {{\cal M}}_{H^+ H^-
A}=l_h \cbma \lambda_{H^+H^-h}+l_H \sbma \lambda_{H^+H^-H}
\eeqn

Cross sections here are very small here. For instance for the set
of parameters considered above and with $\sqrt{s}=2$TeV, $AAA$
production is about $1.4 10^{-7}$fb! Although a full study
allowing a much larger parameter range (including  $M_A$) is in
order, it seems that a $\lambda_b$ measurement would be out of
reach.

\subsection{Stage 5}
As we have seen earlier (see Eq.~\ref{compact}) the effect of the
third combination of parameters, $\lambda_c$, can only be observed
in processes involving a vertex with 4 Higgses. The first
threshold where such a vertex contributes is a $Zhhh$ final state,
which we could have classified in stage 2 (with a $ZHh$ final
state). However even for a \sm Higgs $Zhhh$ or $\nu_e \bar{\nu}_e
hhh$ at a 10TeV LC with a luminosity as high as
$10^{35}$cm$^{-2}$s$^{-1}$ yields only about 5 events per
year\cite{hhhbattagliaboos}! Thus the prospect for a useful
measurement looks grim especially that in $hhhh$ the $\lambda_c$
effect is screened as $\cbma^4$! Quartic couplings where this
contribution is not screened involve any combination of the heavy
Higgses ($H^\pm, H,A$). Triple $H$ production $ZHHH$ is  not
operative since it is triggered by $ZH$ production while quadruple
production of the heavy Higgses are too tiny to be exploited. Thus
a full reconstruction may prove to be impossible if the full set
$\lambda_{1-7}$ is present.

\setcounter{equation}{0}
\section{Effects from higher order operators}
Up to now we have only discussed the effects of the dim-4
operators. Higher order operators are doomed to contribute less
significantly, as their effects are explicitly screened by a high
scale. We will illustrate this case by considering only three new
operators, and restrict ourselves to a few Higgs self couplings to
make the point. We consider
\beqn
V_{eff} \ra V_{eff}+\frac{1}{\Lambda^2} \left\{ \tilde{\kappa}_1
(H_1 H_1^*)^3 + \tilde{\kappa}_2 (H_2 H_2^*)^3 +\tilde{\kappa}_3
(H_1 H_1^*)^2 (H_2 H_2^*) \right\}
\eeqn

For notational ease we will use
\beq
\kappa_i= \frac{v^2}{\Lambda^2} \tilde{\kappa}_i
\eeq
with $\Lambda$ the scale of new physics.  We find
\beqn
\label{hhhkappa}
\lambda_{hhh}^\kappa&=&-\frac{e^2}{\stw^2} \sbpa \cta \nonumber \\
&+&(\lone -6 \kappa_1 \cb^2 -2 \kappa_3 \sb^2) \; \cb \sa^3-
(\ltwo-6 \kappa_2 \sb^2) \; \ca^3 \sb+ \frac{1}{2}
 ((\lthree +\lfour -2 \kappa_3 \cb^2)+\lfive)\; \sta \cbpa \nonumber \\
 &  +& \lsix \sa^2 (\cbpa +2 \ca \cb) + \lseven \ca^2 (\cbpa -2\sa
 \sb)\nonumber \\
 & -& 4\kappa_1 \cb^3 \sa^3 +4 \kappa_2 \sb^3 \ca^3
+ 4 \sb \cb^2 \ca \sa^2 \kappa_3
\eeqn

Note that we have split the effect of the new contributions in two
parts. The first (second line of Eq.~\ref{hhhkappa}) can be viewed
as  a shift in $\lambda_{1,2,3}$ while  the other (last line in
Eq.~\ref{hhhkappa}) can be considered as a genuine new
contribution beyond the effects of the dim-4 operators. The shifts
mean that the combinations $(\lone -6 \kappa_1 \cb^2 -2 \kappa_3
\sb^2)$,$(\ltwo-6 \kappa_2 \sb^2)$ and $(\lthree -2 \kappa_3
\cb^2)$ replace $\lambda_{1,2,3}$, respectively, in the definition
of $\alpha$, $m_{h,H}$ in Eqs.~\ref{tanadef}-\ref{mhdef}. Again,
this means that even in the absence of any dim-4 operator, the
dim-6 operators as defined above will also affect the Higgs masses
and couplings to fermions and vector bosons. Moving to the mass
basis, keeping as extra parameters $\lambda_{5,6,7}$ and
$\kappa_{1,2,3}$ we get

\beqn
 \lambda_{hhh}^\kappa= \lambda_{hhh}\;\; -4\kappa_1 \cb^3 \sa^3 +4 \kappa_2 \sb^3 \ca^3
+ 4 \sb \cb^2 \ca \sa^2 \kappa_3
\eeqn

and

\beqn
 \lambda_{Hhh}^\kappa= \lambda_{Hhh}
- 4 \left( \cb^3 \sa^2 \ca \kappa_1 + \sb^3 \ca^2 \sa \kappa_2 -
 (\frac{2}{3}-\sa^2) \cb^2 \sb \sa \kappa_3
 \right)
\eeqn

Where  $\lambda_{H/h\;hh}$  are given in
Eqs.~\ref{lhhh}-\ref{lHhh}.

We see that the higher order operators are not further reduced by
the decoupling factor $\cbma$ and that all $\kappa_i$ contribute
to all the self-couplings, unlike with $\lambda_i$ where we are
only left with a combination of two couplings. This means that if
one ideally have measured all the masses and couplings to ordinary
fermions and quite precisely all the trilinear Higgs
self-couplings one could tell whether higher order operators are
contributing. However considering the foreseen precision on the
extraction of the Higgs self-couplings and the expected small
contribution of the higher order terms, this would seem to be
overly optimistic.

\setcounter{equation}{0}
\section{Conclusion}
A dedicated study of  double Higgs production at a high luminosity
LC\cite{Teslaphysics,hhhbattagliaboos} within the \sm has shown
that it is very difficult to extract the Higgs self-couplings with
a precision better than $20\%$ in the first stage of a LC
improving to slightly better than $10\%$\cite{hhhbattagliaboos} at
a multi-TeV LC facility, even in the most favourable case of a
Higgs light enough that decays into $b\bar b$. In the decoupling
limit, the lightest SUSY Higgs will have properties very similar
to that of the \sm and thus we would also get a precision on its
self-coupling with a very similar precision. Unfortunately we have
shown that in this limit once we have measured the mass of the
Higgs (which will be known at better than the per-mil level) and
its couplings to the ordinary \sm particles (with a precision of a
few per-cent), the precision attained in the self-couplings will
not be sufficient to reveal new physics. Indeed effects from
``anomalous" operators affecting the Higgs potential have a direct
impact on the Higgs mass and the couplings to fermions. When these
are taken into account additional effects in the self-couplings
are screened either by mixing angles (dim-4 operators) or large
scales. Even if we are not in the decoupling regime and even if we
restrict oneself to the leading dim-4 operators, we have shown
that measurements of all the possible trilinear self-couplings
would not allow to reconstruct the most general lowest dimension
Higgs potential. To achieve this one needs to measure some of the
quartic couplings. However an analysis that would take into
account the measurements of the Higgs masses and their couplings
to the ordinary particles should give some useful constraints. An
analysis along these lines completed with the extraction of some
of the trilinear self-couplings at different stages of the LC is
under way\cite{wehhhexp}.

{\bf Acknowledgement}

We thank G.~B\'elanger for a careful reading of the manuscript and
helpful comments. We also acknowledge useful discussions with
M.~Battaglia, P.~N.~Pandita and M.~Dubinin. A.S. is partially
supported by a CERN-INTAS grant No 99-0377 and a RFFR grant No
01-02-16710.

\end{document}